\documentclass[%
reprint,
amsmath,amssymb,
prb,
]{revtex4-1}

\usepackage{graphicx}
\usepackage{dcolumn}
\usepackage{bm}
\usepackage{bbold}
\usepackage{feynmp}									
\usepackage{color}

\begin{document}
\setlength{\unitlength}{1mm}		
\begin{fmffile}{fgraphs}				


\title{Keldysh functional renormalization group for electronic properties of graphene}

\author{Christian Fr\"a\ss dorf}
\affiliation{%
Dahlem Center for Complex Quantum Systems and, Institut f\"ur Theoretische Physik, \\Freie Universit\"at Berlin, Arnimallee 14, 14195 Berlin, Germany}%

\author{Johannes E. M. Mosig}%
\affiliation{%
Department of Mathematics and Statistics, University of Otago, PO Box 56, Dunedin 9054, New Zealand}%

\date{\today}%

\begin{abstract}
We construct a nonperturbative nonequilibrium theory for graphene electrons interacting via the instantaneous Coulomb interaction by combining the functional renormalization group method with the nonequilibrium Keldysh formalism. The Coulomb interaction is partially bosonized in the forward scattering channel resulting in a coupled Fermi-Bose theory. Quantum kinetic equations for the Dirac fermions and the Hubbard-Stratonovich boson are derived in Keldysh basis, together with the exact flow equation for the effective action and the hierarchy of one-particle irreducible vertex functions, taking into account a possible non-zero expectation value of the bosonic field. Eventually, the system of equations is solved approximately under thermal equilibrium conditions at finite temperature, providing results for the renormalized Fermi velocity and the static dielectric function, which extends the zero-temperature results of Bauer~\textit{et~al.}, Phys.\ Rev.\ B {\bf 92}, 121409 (2015).

\begin{description}

\item[PACS numbers]
11.10.Hi, 71.10.-w, 72.10.Bg, 72.80.Vp, 73.22.Pr, 73.61.-r, 81.05.ue

\end{description}

\end{abstract}

\pacs{Valid PACS appear here}
\maketitle


\section{Introduction}
\label{sec:Introduction}
The band structure of graphene features two isolated points where valence and conduction bands touch.\cite{CastroNetoGuineaNovoselovGeim2009, Dassarma2011, Peres2010} At these touching points the electrons have a linear energy-momentum dispersion, similar to massless relativistic Dirac particles.\cite{Wallace1947} This pseudorelativistic band structure is responsible for the appearance of phenomena usually related to the relativistic domain, such as Klein tunneling through potential barriers,\cite{Klein1929, Cheianov2006, Katsnelson2006a, Beenakker2008} the Zitterbewegung,\cite{Katsnelson2006b} or an anomalous quantized Hall effect.\cite{GusyninSharapov2005, Novoselov2005, Zhang2005, Novoselov2007}

For a description of realistic graphene samples, effects of disorder and electron-electron interactions  have to be added to this idealized band structure. Disorder smears out the singularity at the nodal point, but preserves many of graphene's remarkable electronic properties,\cite{CastroNetoGuineaNovoselovGeim2009, Dassarma2011} and even leads to fundamentally new phenomena by itself, such as the absence of Anderson localization if disorder does not couple the nodal points.\cite{Bardarson2007, Nomura2007, Sanjose2007} The effect of interactions is most pronounced if the singularity in the density of states of the noninteracting theory is not smeared by disorder and the chemical potential is close to the nodal point.\cite{KotovEtal2012} The vanishing carrier density at the nodal point at zero temperature~\cite{HobsonNierenberg1953} implies the absence of screening, which leads to strongly enhanced interaction corrections. In particular, interactions are found to effectively renormalize the Fermi velocity at the nodal point, and the corrections to the velocity diverge logarithmically in the low-temperature limit.\cite{GonzalezGuineaVozmediano1993, Vozmediano2011} These logarithmic corrections have recently been verified experimentally, and good agreement with theoretical calculations was reported.\cite{EliasEtal2011}

Although there is consensus about the way in which interactions affect the electronic structure of graphene,\cite{KotovEtal2012} a quantitative evaluation of the corrections proved to be problematic. The dimensionless interaction strength for the electrons in graphene is $\alpha = e^2/\epsilon_0 \hbar v_F$, which approximately equals~$2.2$ in the freestanding case in vacuum ($\epsilon_0 = 1$). For such a large interaction strength a perturbative calculation of the renormalization effect cannot be reliable, and at first sight the reported agreement of one-loop perturbation theory with the experimentally observed increase of the Fermi velocity appears surprising. Indeed, a two-loop calculation leads a completely different result, a decrease of the Fermi velocity for small momenta.\cite{Mishchenko2007, VafekCase2008, BarnesEtal2015} An alternative approach is to make use of the largeness of the number of fermion species (which is $N_{\rm f} = 4$ in graphene), and a perturbation theory in $1/N_{\rm f}$ gives results largely consistent with the approach based on a perturbative treatment of the interaction strength.\cite{Foster2008, Gonzalez1999}

To address such a situation in which no small parameter, to organize a perturbative expansion, is available, nonperturbative methods have been applied to the problem of interacting Dirac fermions in two dimensions. One of those nonperturbative methods is the functional renormalization group (fRG), which shares some features with the celebrated Wilsonian renormalization group,\cite{AltlandSimonsBook, BagnulsBervillier2001} but rigorously extends the concept of flowing coupling constants to (one-particle irreducible vertex) functions. Initiated by Wetterich,\cite{Wetterich1991, Note1}
this method has found widespread applications in high energy and in condensed matter physics.\cite{MetznerHonercampEtal2012, BergesTetradisWetterich2002, Wetterich2001, GiesWetterich2002, KopietzBook} Of particular relevance to the present problem is the work of Bauer \textit{et~al.},\cite{BauerKopietz2015} who studied the Fermi velocity renormalization and the static dielectric function in graphene at zero temperature using the fRG framework and found excellent agreement with the experiment, surpassing the results of the conventional perturbative methods.

As powerful as the fRG is, it clearly has its limitations when used within its most commonly employed formulation in imaginary time. First and foremost, true nonequilibrium phenemena (beyond linear response) are out of reach of the Matsubara formalism. Second, even for linear response calculations the imaginary time formalism requires an analytical continuation from imaginary to real time at the end of a calculation, which may pose technical difficulties. The appropriate framework to describe true nonequilibrium dynamics is the Keldysh formalism.\cite{ChouSuYu1985, KamenevBook, RammerBook} The Keldysh formalism has the additional advantage that it erases the necessity of analytical continuations, which may also makes it a useful tool for equilibrium applications. Gezzi \textit{et al.}\ implemented a Keldysh formulation of fRG for applications to impurity problems.\cite{GezziPruschkeMeden2006} Jakobs \textit{et al.}\ further developed the theory, constructing a ``Keldysh-compatible'' cutoff scheme that respects causality, with applications to quantum dots and nanowires coupled to external baths.\cite{JakobsMedenSchoeller2007, JakobsPletyukovSchoeller2010} Keldysh formulations of fRG were also developed for various systems involving bosons.\cite{BergesMesterhazy2012, BergesHoffmeister2009, KlossKopietz2011, GasenzerPawlowski2008, GasenzerKesslerPawlowski2010}

In the present article we construct a Keldysh fRG theory for interacting Dirac fermions, as they occur at the nodal points in the graphene band structure. As a test of the formalism, we recalculate the Fermi velocity renormalization and the static dielectric function in graphene, finding full agreement with the zero-temperature Matsubara-formalism calculation of Bauer \textit{et al.}~\cite{BauerKopietz2015} We also extend the calculation to finite temperatures, an extension that in principle is possible within the Matsubara formalism, too, but that comes at no additional calculational cost when done in the Keldysh formalism. We leave applications to true nonequilibrium properties of graphene for future work, but already notice that there is a vast body of perturbative (or in other ways approximate) true nonequilibrium theoretical results for graphene that such a theory can be compared with, see, \textit{e.g.}, Refs.~\onlinecite{GorbarGusynin2002, GusyninSharapovCarbotte2007, SchuettGornyiMirlin2011, NarozhnySchuettMirlin2015}. Although our theory focuses on graphene, a major part of the formalism we develop here is also applicable to conventional nonrelativistic fermions. 

The extension of an imaginary-time fRG formulation to a Keldysh-based formulation involves quite a number of subtle steps and manipulations. One issue is the choice of a cut off scheme, which preferentially is compatible with the causality structure of the Keldysh formalism and, for equilibrium applications, with the fluctuation-dissipation theorem.\cite{JakobsMedenSchoeller2007, JakobsPletyukovSchoeller2010} Another issue is the possibility of an arbitrary nonequilibrium initial condition and the truncation of the (in principle) infinite hierarchy of flow equations in the fRG approach. To do justice to these issues, we have chosen to make this article self contained, although we tried to keep the discussion of standard issues as brief as possible.

The outline of the paper is as follows: In Sec.~\ref{sec:NonequilibriumQuantumFieldTheory} we introduce the formal aspects of nonequilibrium quantum field theory, using the Keldysh technique applied to graphene. The originally purely fermionic problem is formulated as a coupled fermion-boson problem by means of a Hubbard-Stratonovich transformation, singling out the dominant interaction channel. The ideas of the functional renormalization group are reviewed in Sec.~\ref{sec:NonequilibriumFunctionalRenormalizationGroup}, where we combine them with the nonequilibrium Keldysh formalism. We implement an infrared regularization and derive the exact spectral Dyson equations and quantum kinetic equations, as well as an exact flow equation, which incorporates all of the nonperturbative aspects of the theory. Finally, we perform a vertex expansion leading to an exact, infinite hierarchy of coupled integro-differential equations for the one-particle irreducible vertex functions. Section~\ref{sec:ThermalEquilibrium} deals with a solution of our theory in thermal equilibrium. We discuss the necessary limitations for the construction of suitable regulator functions, which preserve causality and, at the same time, the fluctuation-dissipation theorem, allowing a solution of the quantum kinetic equations at all scales. We further present a simple truncation scheme for the calculation of the Fermi velocity and static dielectric function at finite temperature, extending the results of Bauer \textit{et al.}~\cite{BauerKopietz2015}

\section{Nonequilibrium Quantum Field Theory}
\label{sec:NonequilibriumQuantumFieldTheory}
This section mainly serves as an introduction to the Fermi-Bose quantum field theory of interacting electrons in graphene in the nonequilibrium Keldysh formulation. The reader who is familiar with this formulation may skim through our notational conventions and continue reading at section \ref{sec:NonequilibriumFunctionalRenormalizationGroup}.

We consider interacting Dirac fermions in two dimensions, which are described by a grand canonical Hamiltonian in the Heisenberg picture
\begin{align}
H(t) = H_{\textrm{f}}(t) + H_{\textrm{int}}(t) \,.
\label{eq:DecompositionInteractingHamiltonian}
\end{align}
Here $H_{\textrm{f}}$ describes the low energy approximation of free electrons hopping on the honeycomb lattice, and $H_{\textrm{int}}$ contains the interaction effects. The first term reads~\cite{GusyninSharapovCarbotte2007} ($\hbar = c = 1$)
\begin{align}
H_{\textrm{f}}(t) &= \int_{\vec{r}} \Psi^{\dagger}(\vec{r}, t) \big( - \mu + e \varphi(\vec{r}, t) \big) \Psi(\vec{r}, t)
\label{eq:FreeHamiltonian} \\
& - i v_F \int_{\vec{r}} \Psi^{\dagger}(\vec{r}, t) \sigma_0^s \otimes \vec{\Sigma} \cdot \left( \nabla + ie \vec{A}(\vec{r}, t) \right) \Psi(\vec{r}, t) \,, \nonumber
\end{align}
with the chemical potential $\mu$ and the external electromagnetic potentials $\varphi$ and $\vec{A}$. The Dirac electrons are described by eight-dimensional spinors, where we choose the basis as $\Psi \equiv \begin{pmatrix} \Psi_{\uparrow} & \Psi_{\downarrow} \end{pmatrix}^{\intercal}$, with 
\begin{equation}
\Psi_{\sigma} \equiv \begin{pmatrix} \psi_{A K_+} & \psi_{B K_+} & \psi_{B K_-} & \psi_{A K_-} \end{pmatrix}_{\sigma}^{\intercal} \,.
\label{eq:BasisPsi}
\end{equation}
The indices $\sigma = \uparrow, \downarrow$ denote the spin, $K_{\pm}$ the valley- and $A/B$ the sublattice degree of freedom. Further, $\sigma_0^s$ is the two-dimensional unit matrix acting in spin space and $\Sigma_{x, y} = \tau_3 \otimes \sigma_{x, y}$, with the Pauli matrices $\tau_3$ and $\sigma_{x, y}$ acting in valley and sublattice space, respectively. The interaction part is given by the instantaneous Coulomb interaction
\begin{equation}
H_{\textrm{int}}(t) = \frac{1}{2} \int_{\vec{r}, \vec{r}'} \delta n(\vec{r}, t) V(\vec{r} - \vec{r}') \delta n(\vec{r}', t) \,, 
\label{eq:InteractionHamiltonian}
\end{equation}
where
\begin{align}
V(\vec{r} - \vec{r}') &= \frac{e^2}{\epsilon_0 |\vec{r} - \vec{r}'|} \,, \\
\delta n(\vec{r}, t) &= \Psi^{\dagger}(\vec{r}, t) \Psi(\vec{r}, t) - \tilde{n}(\vec{r}, t) \,,
\label{eq:DefinitionDeltaNRealTime}
\end{align}
and $\epsilon_0$ is the dielectric constant of the medium, being unity for freestanding graphene in vaccuum. Here the term $\tilde{n}(\vec{r}, t)$ is a background charge density, representing the charge accumulated on a nearby metal gate. Away from the charge neutrality point it essentially acts as a counterterm, which removes the zero wavenumber singularity of the Coulomb interaction at finite charge carrier density.

\subsection{Single-particle Green functions}
\label{sec:GreenFunctions}
\begin{figure}
	\centering
		\includegraphics[width=.80\columnwidth]{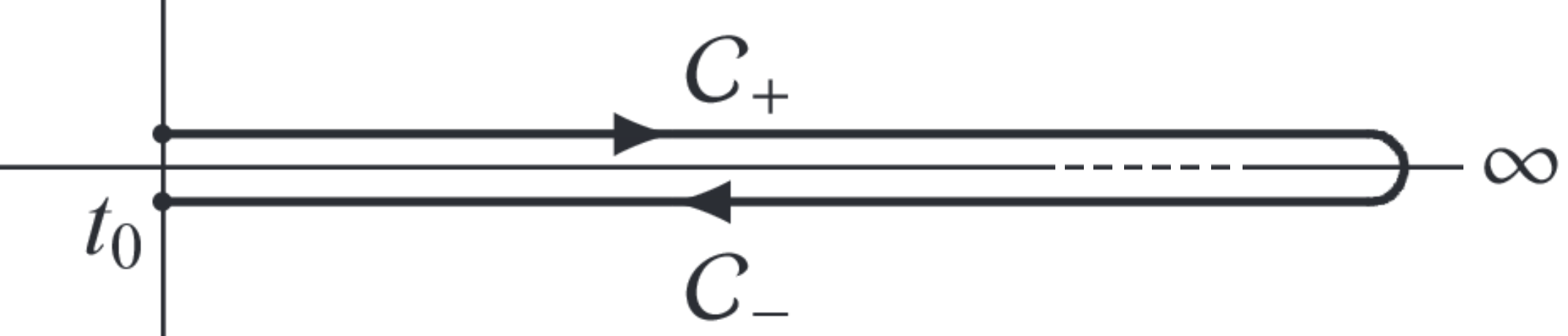}
	\caption{Schwinger-Keldysh time contour in the complex time plane with reference time $t_0$ as starting and end point. $\mathcal{C}_+$ and $\mathcal{C}_-$ are the forward and backward branch, respectively.}
	\label{fig:TimeContour}
\end{figure}

Relevant physical observables can be expressed as correlation functions of the field operators, and the purpose of a field-theoretic treatment is to provide a formalism in which such correlation functions can be calculated efficiently. For an explicitly time-dependent Hamiltonian, such as the one above, one considers the evolution of the field operators along the ``Schwinger-Keldysh contour'',\cite{ChouSuYu1985, RammerBook, KamenevBook} a closed time contour starting at a reference time $t_0$, extending to $+ \infty$, and eventually returning from $+ \infty$ to $t_0$, see Fig.~\ref{fig:TimeContour}.
Consequently, the time arguments $t$ of the field operators are elevated to the ``contour time'', and the building blocks of the theory are formed by the expectation values of ``path ordered'' products of the field operators. The concept of path ordering generalizes the concept of (imaginary) time ordering, such that field operators with a higher contour time appear to the right of operators with a lower contour time. In particular, the single-particle propagator reads
\begin{align}
G_{ij}^{T_{\mathcal{C}}}(\vec{r}, t;\vec{r}', t') &= - i \langle T_{\mathcal{C}} \psi_i(\vec{r}, t) \psi_j^{\dagger}(\vec{r}', t') \rangle \,,
\label{eq:ContourTimeGreenFunction}
\end{align}
where the indices $i,j$ represent collectively the sublattice, valley and spin degrees of freedom. $T_{\mathcal{C}}$ is the contour-time ordering operator and the expectation value is performed with respect to some initial density matrix given at a reference time $t_0$
\begin{equation}
\langle \cdots \rangle = \textrm{Tr}[\rho(t_0) \cdots] \,.
\label{eq:DefinitionAverage}
\end{equation}
Since there are four possibilities where the two time variables can be located to each other with respect to the two time branches $\mathcal{C}_+$ and $\mathcal{C}_-$, one can map the contour-ordered Green function to a $2 \times 2$ matrix representation with time-arguments defined on the real axis
\begin{align}
\boldsymbol{G}_{ij}(\vec{r}, t;\vec{r}', t') &= \begin{pmatrix} G_{ij}^{++}(\vec{r}, t;\vec{r}', t') & G_{ij}^{+-}(\vec{r}, t;\vec{r}', t') \\[1mm] G_{ij}^{-+}(\vec{r}, t;\vec{r}', t') & G_{ij}^{--}(\vec{r}, t;\vec{r}', t') \end{pmatrix} \nonumber \\
&= \begin{pmatrix} G_{ij}^{T}(\vec{r}, t;\vec{r}', t') & G_{ij}^{<}(\vec{r}, t;\vec{r}', t') \\[1mm] G_{ij}^{>}(\vec{r}, t;\vec{r}', t') & G_{ij}^{\bar{T}}(\vec{r}, t;\vec{r}', t') \end{pmatrix} \,.
\label{eq:MatrixGreenFunction}
\end{align}
The constituents of this matrix are the time ordered, anti-time ordered, greater and lesser Green function, respectively,
\begin{subequations}
\begin{align}
G_{ij}^{T}(\vec{r}, t;\vec{r}', t') &= - i \langle T \psi_i(\vec{r}, t) \psi_j^{\dagger}(\vec{r}', t') \rangle \,, \\
G_{ij}^{\bar{T}}(\vec{r}, t;\vec{r}', t') &= - i \langle \bar{T} \psi_i(\vec{r}, t) \psi_j^{\dagger}(\vec{r}', t') \rangle \,, \\
G_{ij}^{>}(\vec{r}, t;\vec{r}', t') &= - i \langle \psi_i(\vec{r}, t) \psi_j^{\dagger}(\vec{r}', t') \rangle \,, \\
G_{ij}^{<}(\vec{r}, t;\vec{r}', t') &= + i \langle \psi_j^{\dagger}(\vec{r}', t') \psi_i(\vec{r}, t) \rangle \,.
\label{eq:DefinitionComponentsOfMatrixGreenFunction}
\end{align}
\end{subequations}
By definition, these functions are linearly dependent and subject to the following constraint,\cite{ChouSuYu1985, RammerBook, KamenevBook}
\begin{equation}
\hat{G}^T - \hat{G}^< - \hat{G}^> + \hat{G}^{\bar{T}} = 0 \,,
\label{eq:GFConstraint}
\end{equation}
which allows a basis transformation to three linearly independent propagators. This transformation is given by the involutional matrix $\tau_1 L$, where $\tau_1$ is a Pauli matrix and $L$ is the orthogonal matrix 
\begin{equation}
L = \frac{1}{\sqrt{2}} \begin{pmatrix} 1 & - 1 \\ 1 & 1 \end{pmatrix} \,,
\label{eq:OriginalKeldyshRotationMatrix}
\end{equation}
originally introduced by Keldysh.\cite{Keldysh1964} Its application to Eq.~\eqref{eq:MatrixGreenFunction} yields
\begin{equation}
\tau_1 L \boldsymbol{\hat{G}} (\tau_1 L)^{-1} = \begin{pmatrix} \hat{G}^K & \hat{G}^R \\ \hat{G}^A & 0 \end{pmatrix} \,,
\label{eq:RotatedMatrixGreenFunction}
\end{equation}
with
\begin{subequations}
\begin{align}
\hat{G}^R &= \tfrac{1}{2} \left( \hat{G}^T - \hat{G}^< + \hat{G}^> - \hat{G}^{\bar{T}} \right) \,, \\
\hat{G}^A &= \tfrac{1}{2} \left( \hat{G}^T + \hat{G}^< - \hat{G}^> - \hat{G}^{\bar{T}} \right) \,, \\
\hat{G}^K &= \tfrac{1}{2} \left( \hat{G}^T + \hat{G}^< + \hat{G}^> + \hat{G}^{\bar{T}} \right) \,.
\label{eq:RAKGreenFunctions}
\end{align}
\end{subequations}
The functions $\hat{G}^{R/A/K}$ are the retarded, advanced and Keldysh propagators, respectively. The latter one is also known as the statistical propagator. They obey the symmetry relations
\begin{equation}
(\hat{G}^R)^{\dagger} = \hat{G}^A \,, \quad (\hat{G}^K)^{\dagger} = - \hat{G}^K \,,
\label{eq:SymmetryOfRAKGFs}
\end{equation}
as well as the causality relations~\cite{ChouSuYu1985, RammerBook, KamenevBook}
\begin{subequations}
\begin{align}
\hat{G}^R(\vec{r}, t; \vec{r}', t') = 0 \,, \quad \textrm{if} \quad t < t' \,, \\
\hat{G}^A(\vec{r}, t; \vec{r}', t') = 0 \,, \quad \textrm{if} \quad t > t' \,.
\label{eq:CausalityRelations}
\end{align}
\end{subequations}

Explicit expressions for the free propagators may easily be obtained in thermal equilibrium and in the absence of the electromagnetic potentials. To this end we send the reference time $t_0 \rightarrow -\infty$ and Fourier transform the field operators following the conventions
%
\begin{align}
\psi_i(\vec{r}, t) &= \int_{\vec{k}, \varepsilon} e^{+ i \vec{k} \cdot \vec{r} - i \varepsilon t} \psi_i(\vec{k}, \varepsilon) \,,
\label{eq:FourierTransformFieldOperators}
\end{align}
%
with $\int_{\vec{k}, \varepsilon} \equiv \int \frac{d^2 k}{(2 \pi)^2} \frac{d \varepsilon}{2 \pi}$. After a short calculation one finds
\begin{subequations}
\begin{align}
\!\! \hat{G}_0^{R/A}(\vec{k}, \varepsilon) &= \frac{1}{\sigma_0^s \otimes \left(\Sigma_0 (\varepsilon + \mu \pm i 0) - v_F \vec{\Sigma} \cdot \vec{k} \right)} \,, \\
\hat{G}_0^K(\vec{k}, \varepsilon) &= \textrm{tanh} \left(\frac{\varepsilon}{2T} \right) \left( \hat{G}_0^R(\vec{k}, \varepsilon) - \hat{G}_0^A(\vec{k}, \varepsilon) \right) \,,
\label{eq:RAKGreenFunctionsNoninteractingGraphene}
\end{align}
\end{subequations}
where $\Sigma_0 = \tau_0 \otimes \sigma_0$ is the $4 \times 4$ unit matrix in valley-sublattice space. Note that the entire statistical information of the system is contained in the Keldysh propagator. These expressions may be further simplified by expanding the propagators in the chiral basis
\begin{equation}
\hat{G}_0^{R/A}(\vec{k}, \varepsilon) = \sum_{\pm} \hat{\mathcal{P}}_{\pm}(\hat{k}) G_{\pm, 0}^{R/A}(k, \varepsilon) \,,
\label{eq:NoninteractingRAGreenFunctionsChiralDecomposition}
\end{equation}
in which $\mathcal{\hat{P}}_{\pm}(\hat{k})$ are the chiral projection operators
\begin{equation}
\hat{\mathcal{P}}_{\pm}(\hat{k}) = \sigma_0^s \otimes \left( \frac{\Sigma_0 \pm \vec{\Sigma} \cdot \hat{k}}{2} \right) \,,
\label{eq:ChiralProjectors}
\end{equation}
with $\hat{k} = \vec{k}/k$. In the chiral basis the propagators then take the simple form
\begin{subequations}
\begin{align}
G_{\pm, 0}^R(k, \varepsilon) &= \frac{1}{(\varepsilon + \mu + i 0) \mp v_F k} \,, \\
G_{\pm, 0}^A(k, \varepsilon) &= \frac{1}{(\varepsilon + \mu - i 0) \mp v_F k} \,, \\
G_{\pm, 0}^K(k, \varepsilon) &= - 2 \pi i \, \textrm{tanh} \left(\frac{\varepsilon}{2T} \right) \delta(\varepsilon + \mu \mp v_F k) \,.
\label{eq:NoninteractingRAGreenFunctionsInChiralBasis}
\end{align}
\end{subequations}

The density of electrons in the system is given by
\begin{align}
n(\vec{r}, t) &= - i \, \textrm{tr} \, \hat{G}^<(\vec{r}, t, \vec{r}, t) \nonumber \\
&= - \frac{i}{2} \textrm{tr} \Big( \hat{G}^K - \big( \hat{G}^R - \hat{G}^A \big) \Big)(\vec{r}, t, \vec{r}, t) \,,
\label{eq:ParticleDensity}
\end{align}
which is formally divergent. The charge carrier density, however, which is defined as~\cite{GusyninSharapovCarbotte2007}
\begin{equation}
\bar{n}(\vec{r}, t) = - \frac{i}{2} \textrm{tr} \, \hat{G}^K(\vec{r}, t, \vec{r}, t) \,,
\label{eq:ParticleImbalance}
\end{equation}
is finite. It is a function of the external doping $\mu$ and of the gauge invariant external electromagnetic fields. In the absence of such external fields, it vanishes at the charge neutrality point ($\mu = 0$).

\subsection{Contour-time generating functional}
\label{sec:ContourTimeGeneratingFunctional}
The entire physical content of the theory can be conveniently expressed by the partition function~\cite{ChouSuYu1985, NegeleOrlandBook, KamenevBook, BergesMesterhazy2012, CalzettaHu1988}
\begin{equation}
Z[\eta; \rho] = \left\langle T_{\mathcal{C}} e^{i \eta^{\dagger} \Psi + i \Psi^{\dagger}} \right\rangle \,,
\label{eq:OperatorDefinitionContourTimeGeneratingFunctional}
\end{equation}
which is a generating functional for all $n$-point correlation functions, including the single-particle propagators described above. Its arguments $\eta$ and $\eta^{\dagger}$, where only the former is shown on the left hand side for brevity, are eight component spinorial external source terms. Here and in the remainder of this article we employed a condensed vector notation
\begin{equation}
\eta^{\dagger} \Psi \equiv \int_{\mathcal{C}, x} \eta^{\dagger}(x) \Psi(x) \,, \quad \Psi^{\dagger} \eta \equiv \int_{\mathcal{C}, x} \Psi^{\dagger}(x) \eta(x) \,,
\label{eq:NotationSourceTerms}
\end{equation}
where $x = (\vec{r}, t)$ labels space and (contour-) time coordinates, such that
\begin{equation}
\int_{\mathcal{C}, x} \equiv \int_{\mathcal{C}} dt \int d^2r \,.
\label{eq:NotationContourIntegration}
\end{equation}
The symbol $\mathcal{C}$ indicates that the time integration has to be performed along the Schwinger-Keldysh closed time contour. An important property of the partition function is that it is normalized to unity when the sources are set equal to zero\cite{Note2}
%
\begin{equation}
Z[0; \rho] = \textrm{Tr} \, \rho(t_0) = 1 \,.
\label{eq:NormalizationGeneratingFunctional}
\end{equation}
In fact, this normalization is the very reason for the algebraic identity \eqref{eq:GFConstraint} and it leads to similar constraints for higher order correlation functions, see Ref.~\onlinecite{ChouSuYu1985}. It further ensures that any correlation function computed from the partition function~\eqref{eq:OperatorDefinitionContourTimeGeneratingFunctional} does not contain disconnected bubble diagrams.

The partition function \eqref{eq:OperatorDefinitionContourTimeGeneratingFunctional} can be represented in terms of a fermionic coherent state functional integral as~\cite{NegeleOrlandBook, KamenevBook, ChouSuYu1985, CalzettaHu1988}
\begin{align}
Z[\eta; \rho] = \int \mathcal{D}\psi \mathcal{D}\psi^{\dagger} e^{iS[\psi] + iK_{\rho}[\psi] + i\vec{\eta}^{\dagger} \Psi + i\vec{\Psi}^{\dagger} \eta} \,.
\label{eq:FieldDefinitionContourTimeGeneratingFunctional}
\end{align}
Here $S[\psi]$ is the contour-time action of the system and $K_{\rho}[\psi]$ is the correlation functional, which incorporates the statistical information of the initial density matrix.\cite{CalzettaHu1988, ChouSuYu1985, BergesCox2000} Their dependence on the Grassmann-valued spinor fields $\Psi$ and $\Psi^{\dagger}$ has been abbreviated by $\psi$, as we did for the source field dependence of the partition function. 

The action can be written as a contour-time integral over the Lagrangian $L(t)$
\begin{equation}
S[\psi] = \int_{\mathcal{C}, t} L(t) \,,
\label{eq:Action}
\end{equation}
with
\begin{equation}
L(t) = \int_{\vec{r}} \Psi^{\dagger}(x) i \partial_t \Psi(x) - H(t) \,.
\label{eq:Lagrangian}
\end{equation}
Similarly to the Hamiltonian~\eqref{eq:DecompositionInteractingHamiltonian}, the action decomposes into free contribution and an interaction term,
\begin{equation}
S[\psi] = S_{\textrm{f}}[\psi] + S_{\textrm{int}}[\psi] \,,
\label{eq:DecompositionAction}
\end{equation} 
expressions for which can be obtained immediately by substitution of Eqs.~\eqref{eq:FreeHamiltonian} and \eqref{eq:InteractionHamiltonian}.

The functional $K_{\rho}[\psi]$ describes the initial correlations of the system, corresponding to the density matrix $\rho(t_0)$. It may be expanded in powers of fields as
\begin{widetext}
\begin{align}
K_{\rho}[\psi] = \sum_{m = 0}^{\infty} \frac{(-1)^m}{(m!)^2} \int_{\mathcal{C}, x_m x'_m} \sum_{i_m, i'_m} &K_{\rho}^{(2m)}(x_1 i_1, \ldots, x_m i_m; x'_1 i'_1, \ldots, x'_m i'_m) \nonumber \\ \times &\psi_{i_1}^{\dagger}(x_1) \ldots \psi_{i_m}^{\dagger}(x_m) \psi_{i'_m}(x'_m) \ldots \psi_{i'_1}(x'_1) \,,
\label{eq:InitialCorrelationFunctional}
\end{align}
%
%
\end{widetext}
where the kernels $K_{\rho}^{(2m)}$ are nonvanishing only, if all their respective contour-time arguments equal the initial time $t_0$. The statistical information contained in the kernels $K_{\rho}^{(2m)}$, specifying the correlations present in the initial state, is in a one-to-one correspondence to the statistical information contained in the density matrix.\cite{CalzettaHu1988, BergesCox2000, GasenzerKesslerPawlowski2010} In practice, only a limited set of initial correlations is taken into account, either because of an implicit assumption that the initial state is a thermal equilibrium state for an effectively noninteracting system,\cite{RammerBook, KamenevBook} or as an expression of the finite knowledge that is available about an experimental setup.\cite{BergesCox2000} In the remainder of this work we mainly focus on Gaussian density matrices, \textit{i.e.}, we truncate the series~\eqref{eq:InitialCorrelationFunctional} after the first term, absorbing the statistical information of $K_{\rho}^{(2)}$ into the boundary conditions of the two-point function and simply write $Z[\eta, \rho] \equiv Z[\eta]$. Yet most of our results are not affected by this simplification and valid even in the general case. We come back to this issue in section~\ref{sec:VertexExpansion}, where we comment on some questions regarding the possible implementation of correlated initial states.

Although it is possible to treat the theory presented so far within the formalism of the (fermionic) functional renormalization group,\cite{TanizakiHatsuda2014, MetznerHonercampEtal2012} we here choose a formulation in which a bosonic field is introduced by means of a Hubbard-Stratonovich transformation, that decouples the Coulomb interaction.\cite{NegeleOrlandBook, KamenevBook, KopietzBook} It is well-known that bosonic degrees of freedom, such as Cooper pairs in the celebrated BCS-theory of superconductivity,\cite{NegeleOrlandBook} naturally emerge as collective, low energy degrees of freedom of composite fermions. Therefore, it is reasonable to introduce a collective bosonic field right from the beginning, which captures the dominant contributions of the interaction.

The Hubbard-Stratonovich transformation is an exact integral identity replacing the four-fermion interaction $S_{\textrm{int}}[\psi]$ by a quadratic form of a real bosonic field and a Fermi-Bose interaction
\begin{equation}
e^{i S_{\textrm{int}}[\psi]} = \int \mathcal{D}\phi \, e^{i S_{\textrm{b}}[\phi] + i S_{\textrm{int}}[\psi, \phi]} \,.
\label{eq:HubbardStratonovich}
\end{equation}
The free bosonic part is given by
\begin{equation}
S_{\textrm{b}}[\phi] = \frac{1}{2} \int_{\mathcal{C}, xy} \phi(x) V^{-1}(x - y) \phi(y) \,,
\label{eq:BosonActionQuadraticPart}
\end{equation}
where $V^{-1}$ is the inverse Coulomb interaction, understood in the distributional sense. The interaction term contains a trilinear Yukawa-type interaction and a linear term, describing the coupling of the Hubbard-Stratonovich boson to the background charge density~$\tilde{n}(x)$
\begin{equation}
S_{\textrm{int}}[\psi, \phi] = - \int_{\mathcal{C}, x} \phi(x) \left( \Psi^{\dagger}(x) \Psi(x) - \tilde{n}(x) \right) \,.
\label{eq:FermiBoseCouplingAction}
\end{equation}
Note that the fluctuating Bose field $\phi$ appears on the same footing as the external scalar potential $\varphi$, see Eq.~\eqref{eq:FreeHamiltonian}.

We generalize the Hubbard-Stratonovich transformed partition function by introducing an additional source term $\phi^{\intercal} J$, so that it gives access to bosonic as well as mixed Fermi-Bose correlators. The generalized Fermi-Bose partition function reads
\begin{equation}
Z[\eta, J] = \int \mathcal{D}\psi \mathcal{D}\psi^{\dagger} \mathcal{D}\phi \, e^{i S[\psi, \phi] + i \eta^{\dagger} \Psi + i \Psi^{\dagger} \eta + i \phi^{\intercal} J} \,,
\label{eq:FermiBoseGeneratingFunctional}
\end{equation}
with $S[\psi, \phi] = S_{\textrm{f}}[\psi] + S_{\textrm{b}}[\phi] + S_{\textrm{int}}[\psi, \phi]$. It fulfills the same normalization condition, when the sources are set to zero, as the purely fermionic partition function
\begin{equation}
Z[0, 0] = 1 \,.
\label{eq:NormalizationFermiBoseGeneratingFunctional}
\end{equation}

\subsection{Real-time representation}
\label{sec:RealTimeRepresentation}
Although the contour-time representation allows for a compact and concise notation during any step of a calculation, it is desirable to formulate the theory in a single-valued ``physical'' time which appeals to physical intuition and transparency. Hereto one splits the contour $\mathcal{C}$ into forward ($\mathcal{C}_+$) and backward ($\mathcal{C}_-$) branch, thereby defining a doubled set of fields, $\Psi_{\pm}$ and $\phi_{\pm}$, allocated to the respective branch
\begin{align}
S[\psi, \phi] &= \int_{\mathcal{C}, t} L[\psi, \phi] \nonumber \\
&= \int_{\mathcal{C}_+, t} L[\psi_+, \phi_+] + \int_{\mathcal{C}_-, t} L[\psi_-, \phi_-] \,.
\label{eq:ContourTimeSplitting}
\end{align}
In a next step, one performs a rotation from $\pm$-field space to Keldysh space, using the involutional matrix $\tau_1 L$, see Eq.~\eqref{eq:OriginalKeldyshRotationMatrix}, which was already employed for the rotation of the Green functions in Sec.~\ref{sec:GreenFunctions}. Further, one defines the symmetric and antisymmetric linear combinations of the $\pm$-fields as ``classical'' ($c$) and ``quantum'' ($q$) components, respectively, and combines these into vectors $\boldsymbol{\Psi}, \boldsymbol{\Psi^{\dagger}}$ and $\boldsymbol{\phi}$ as
\begin{align}
\boldsymbol{\Psi} &\equiv \begin{pmatrix} \Psi_c \\ \Psi_q \end{pmatrix} \equiv \tau_1 L \begin{pmatrix} \Psi_+ \\ \Psi_- \end{pmatrix} \,, \quad \boldsymbol{\Psi^{\dagger}} = (\boldsymbol{\Psi})^{\dagger} \,,
\label{eq:DefinitionClassicalQuantumFieldsFermions} \\
\boldsymbol{\phi} &\equiv \begin{pmatrix} \phi_c \\ \phi_q \end{pmatrix} \equiv \frac{1}{\sqrt{2}} \tau_1 L \begin{pmatrix} \phi_+ \\ \phi_- \end{pmatrix} \,.
\label{eq:DefinitionClassicalQuantumFieldsBosons}
\end{align}
%
%
The source fields are rotated and combined into vectors $\boldsymbol{\eta}, \boldsymbol{\eta^{\dagger}}$ and $\boldsymbol{J}$ likewise. Two remarks are in order. First, the mapping of the bosonic source term yields an additional factor of two, due to our choice of normalization in Eq.~\eqref{eq:DefinitionClassicalQuantumFieldsBosons}, which we choose to absorb into a redefinition of $\boldsymbol{J}$. The second remark is concerned about our definition of the Keldysh rotation for the fermionic field $\boldsymbol{\Psi^{\dagger}}$. Some authors prefer a different convention, which was originally proposed by Larkin and Ovchinnikov.\cite{LarkinOvchinnikov1975} In a purely fermionic theory this is reasonable, since it leads to a certain technical simplification. However, this modified rotation is not possible for bosons. In the context of the coupled Fermi-Bose theory we are dealing with, the implementation of the Larkin-Ovchinnikov rotation would lead to an asymmetry in the arising Keldysh structures, which we want to avoid. Therefore, we define the Keldysh rotation as proposed in Eq.~\eqref{eq:DefinitionClassicalQuantumFieldsFermions}. Further, one has to keep in mind that the naming ``classical'' for the fermions is just terminology. For the bosons on the other hand this naming has a physical meaning.

We here summarize the main results of the real-time mapping and explain the structure of the theory obtained after the above Keldysh rotation. For the partition function $Z[\boldsymbol{\eta}, \boldsymbol{J}]$ we find
\begin{align}
Z[\boldsymbol{\eta}, \boldsymbol{J}] = \int \mathcal{D}\boldsymbol{\psi} \mathcal{D}\boldsymbol{\psi^{\dagger}} \mathcal{D}\boldsymbol{\phi} \, e^{i S[\boldsymbol{\psi}, \boldsymbol{\phi}] + i \boldsymbol{\eta^{\dagger}} \tau_1 \boldsymbol{\Psi} + i \boldsymbol{\Psi^{\dagger}} \tau_1 \boldsymbol{\eta} + i \boldsymbol{\phi}^{\intercal} \tau_1 \boldsymbol{J}} \,.
\label{eq:KeldyshFermiBoseGeneratingFunctional}
\end{align}
We have used here the short-hand notation
\begin{align}
\boldsymbol{\eta^{\dagger}} \tau_1 \boldsymbol{\Psi} &\equiv \int_{x} \begin{pmatrix} \eta_c^{\dagger}(x) & \eta_q^{\dagger}(x) \end{pmatrix} \tau_1 \begin{pmatrix} \Psi_c(x) \\ \Psi_q(x) \end{pmatrix} \,,
\label{eq:NotationKeldyshSourceTerm}
\end{align}
in which the Pauli matrix $\tau_1$ acts in Keldysh space, coupling a ``classical'' source to a ``quantum'' field and vice versa. Further, all the time integrations are defined from now on along the forward time branch $\mathcal{C}_+$ only
\begin{equation}
\int_x \equiv \int_{\mathcal{C}_+, x} = \int_{t_0}^{\infty} dt \int d^2r \,.
\label{eq:NotationSingleTimeIntegration}
\end{equation}

The action $S[\boldsymbol{\psi}, \boldsymbol{\phi}]$ is the sum of three contributions,
\begin{equation}
S[\boldsymbol{\psi}, \boldsymbol{\phi}] = S_{\textrm{f}}[\boldsymbol{\psi}] + S_{\textrm{b}}[\boldsymbol{\phi}] + S_{\textrm{int}}[\boldsymbol{\psi}, \boldsymbol{\phi}] \,.
\label{eq:RealTimeAction}
\end{equation}
Its quadratic part in the fermionic sector is given by
\begin{align}
S_{\textrm{f}}[\boldsymbol{\psi}] &= \int_{xy} \begin{pmatrix} \Psi_c^{\dagger}(x) & \Psi_q^{\dagger}(x) \end{pmatrix} \boldsymbol{\hat{G}}_0^{-1}(x, y) \begin{pmatrix} \Psi_c(y) \\ \Psi_q(y) \end{pmatrix} \,.
\label{eq:KeldyshActionQuadraticPartFermions}
\end{align}
The inverse free propagator $\boldsymbol{\hat{G}}_0^{-1}$ has a trigonal matrix structure
\begin{equation}
\boldsymbol{\hat{G}}_0^{-1} = \begin{pmatrix} 0 & (\hat{G}_0^A)^{-1} \\ (\hat{G}_0^R)^{-1} & (\hat{G}_0^{-1})^K \end{pmatrix} \,,
\label{eq:KeldyshInverseFreePropagatorFermions}
\end{equation}
with retarded/advanced $(\hat{G}_0^{R/A})^{-1}$ and Keldysh blocks $(\hat{G}_0^{-1})^K$, which obey the symmetries~\cite{RammerBook, KamenevBook}
\begin{equation}
\left( (\hat{G}_0^R)^{-1} \right)^{\dagger} = (\hat{G}_0^A)^{-1} \,, \quad \left( (\hat{G}_0^{-1})^K \right)^{\dagger} = - (\hat{G}_0^{-1})^K \,.
\label{eq:SymmetriesInverseFreePropagatorsFermions}
\end{equation}
The retarded/advanced blocks are the inverse free retarded/advanced propagators 
\begin{align}
(\hat{G}_0^{R/A})^{-1}(x, y) &= \delta(x - y) \sigma_0^s 
\otimes \left( \Sigma_0 i \mathcal{D}_{y_0} + i v_F \vec{\Sigma} \cdot \mathcal{D}_{\vec{y}} \right) \,,
\label{eq:RetardedAdvancedInverseFreePropagatorFermions}
\end{align}
where the gauge covariant derivative is given by
\begin{equation}
i \mathcal{D}_{x_0} = i \partial_{x_0} \pm i0 + \mu - e \varphi(x) \,, \quad \mathcal{D}_{\vec{x}} = \partial_{\vec{x}} + i e \vec{A}(x) \,.
\label{eq:ContourCovariantDerivatives}
\end{equation}
Note that the regularization term $\pm i0$, which we have written here explicitly, enforces the retarded, respectively advanced, boundary condition. It has to be emphasized that the external gauge fields therein are understood as entirely classical
\begin{subequations}
\begin{align}
\varphi(x) \equiv \varphi_c(x) &= \tfrac{1}{2} \Big( \varphi_+(x) + \varphi_-(x) \Big) \,, \\
\vec{A}(x) \equiv \vec{A}_c(x) &= \tfrac{1}{2} \Big( \vec{A}_+(x) + \vec{A}_-(x) \Big) \,.
\label{eq:ClassicalExternalFieldsDefinition}
\end{align}
\end{subequations}
Since these fields are not quantized, their quantum components in Keldysh space vanish identically. Yet it is formally possible to keep them as source fields, which could be used to generate density-density or current-current correlation functions.\cite{KamenevBook} On the other hand this is not necessary, since we have the single-particle sources $\boldsymbol{\eta}$ at our disposal. In contrast to the retarded and advanced blocks of Eq.~\eqref{eq:KeldyshInverseFreePropagatorFermions}, the Keldysh block $(\hat{G}_0^{-1})^K$ does not take the form of a simple inverse propagator. It carries the statistical information of the theory and can be written as
\begin{equation}
(\hat{G}_0^{-1})^K = - (\hat{G}_0^R)^{-1} \hat{G}_0^K (\hat{G}_0^A)^{-1} \,,
\label{eq:KeldyshComponentInverseFreePropagatorFermions}
\end{equation}
with the noninteracting Keldysh Green function $\hat{G}_0^K$. Since the latter is an anti-hermitian matrix, see Eq.~\eqref{eq:SymmetryOfRAKGFs}, it can be parametrized in terms of a hermitian matrix $\hat{\mathcal{F}}_0$ and the spectral functions $\hat{G}_0^{R/A}$ as~\cite{KamenevBook}
\begin{equation}
\hat{G}_0^K = \hat{G}_0^R \hat{\mathcal{F}}_0 - \hat{\mathcal{F}}_0 \hat{G}_0^A \,.
\label{eq:ParametrizationFreeFermionicKeldyshPropagator}
\end{equation}
Substitution into Eq.~\eqref{eq:KeldyshComponentInverseFreePropagatorFermions} then yields that for noninteracting fermions the Keldysh block of the inverse matrix propagator is a pure regularization term\cite{Note3}
%
\begin{equation}
(\hat{G}_0^{-1})^K = 2 i 0 \hat{\mathcal{F}}_0 \,.
\label{eq:KeldyshComponentInverseFreePropagatorFermionsParametrized}
\end{equation}
Only when interactions are considered the Keldysh block will acquire a finite value. We will come back to this issue in section \ref{sec:DysonAndQuantumKineticEquationsInTheFunctionalRenormalizationGroup}. The free propagator $\boldsymbol{\hat{G}}_0$ is obtained by inverting Eq.~\eqref{eq:KeldyshInverseFreePropagatorFermions}, where the Keldysh structure is given by Eq.~\eqref{eq:RotatedMatrixGreenFunction}.

The quadratic part of the action in the bosonic sector reads
\begin{equation}
S_{\textrm{b}}[\boldsymbol{\phi}] = \frac{1}{2} \int_{xy} \boldsymbol{\phi}^{\intercal}(x) \boldsymbol{D}_0^{-1}(x, y) \boldsymbol{\phi}(y) \,.
\label{eq:KeldyshActionQuadraticPartBoson}
\end{equation}
%
The bosonic matrix $\boldsymbol{D}_0^{-1}$ has the same trigonal structure as the fermionic one
\begin{equation}
\boldsymbol{D}_0^{-1} = \begin{pmatrix} 0 & (D_0^A)^{-1} \\ (D_0^R)^{-1} & (D_0^{-1})^K \end{pmatrix} \,,
\label{eq:KeldyshInverseFreePropagatorBoson}
\end{equation}
with the same symmetry relations as Eq.~\eqref{eq:SymmetriesInverseFreePropagatorsFermions}. Owing to the fact that the bosons are real, the above quantities fulfill the additional symmetries~\cite{KamenevBook, RammerBook}
\begin{equation}
\Big( (D_0^R)^{-1} \Big)^{\intercal} = (D_0^A)^{-1} \,, \quad \Big( (D_0^{-1})^K \Big)^{\intercal} = (D_0^{-1})^K \,.
\label{eq:SymmetriesInverseFreePropagatorsBoson2}
\end{equation}
The retarded and advanced blocks are twice the inverse bare Coulomb interaction
\begin{equation}
(D_0^{R/A})^{-1}(x, y) = 2 V^{-1}(x - y) \,.
\label{eq:RetardedAdvancedInverseFreePropagatorBoson}
\end{equation}
The Keldysh component for bosons has the same structure as the fermionic one
\begin{equation}
(D_0^{-1})^K = - (D_0^R)^{-1} D_0^K (D_0^A)^{-1} \,.
\label{eq:KeldyshComponentInverseFreePropagatorBoson}
\end{equation}
Similarly to the fermionic case we can parametrize the bosonic Keldysh Green function in terms of a hermitian function $\mathcal{B}_0$~\cite{KamenevBook}
\begin{equation}
D_0^K = D_0^R \mathcal{B}_0 - \mathcal{B}_0 D_0^A \,.
\label{eq:ParametrizationFreeBosonicKeldyshPropagator}
\end{equation}
Since the bare Coulomb interaction is instantaneous, the above Keldysh propagator together with the Keldysh block~\eqref{eq:KeldyshComponentInverseFreePropagatorBoson} vanish identically. For that reason we may write
\begin{equation}
\boldsymbol{D}_0^{-1} = 2 \boldsymbol{V}^{-1} \equiv 2 V^{-1} \tau_1 \,.
\label{eq:AlternativeKeldyshInverseFreePropagatorBoson}
\end{equation}
Again, the interaction with the fermions will eventually lead to a finite bosonic Keldysh self-energy and, hence, a nonvanishing Keldysh propagator as in the fermionic case.

Finally we discuss the Fermi-Bose interaction term. Its linear counterterm maps in the same way as the sources do, but with the important difference that the quantum component $\tilde{n}_q(x)$ is identically zero. Nevertheless, we may still use the Keldysh vector notation for this term as well. The trilinear term maps to four interaction terms in real-time, which can be arranged in a matrix form similar to Eq.~\eqref{eq:KeldyshActionQuadraticPartFermions},
\begin{align}
S_{\textrm{int}}[\boldsymbol{\psi}, \boldsymbol{\phi}] = &- \int_x \boldsymbol{\Psi^{\dagger}}(x) \begin{pmatrix} \phi_q(x) & \phi_c(x) \\ \phi_c(x) & \phi_q(x) \end{pmatrix} \boldsymbol{\Psi}(x) \nonumber \\
&+ 2 \int_x \boldsymbol{\phi}^{\intercal}(x) \tau_1 \boldsymbol{\tilde{n}}(x) \,.
\label{eq:KeldyshFermiBoseCoupling}
\end{align}
%
Note the factor of two in front of the linear term in comparison to the linear source term, which could not be absorbed into a redefinition of any of those fields as was the case for $\boldsymbol{J}$. Further observe that the classical components of the fluctuating Bose field appear in the same off-diagonal position as the external gauge field $\varphi$ does in Eq.~\eqref{eq:KeldyshActionQuadraticPartFermions}. The quantum components on the other hand are located in the diagonal.

Now that all of our notational conventions have been established we can move on to the central part of this work.

\section{Nonequilibrium Functional Renormalization Group}
\label{sec:NonequilibriumFunctionalRenormalizationGroup}
The idea of the functional renormalization group is to modify the bare action of the theory by introducing a dependence on a parameter $\Lambda$, in such a way that the partition function can be easily (and exactly) calculated if $\Lambda$ is set equal to an initial value $\Lambda_0$, whereas the true physical system corresponds to $\Lambda = 0$. Using the solution of the modified partition function at $\Lambda = \Lambda_0$, one obtains the ``physical'' partition function at $\Lambda=0$ by tracking its changes upon lowering $\Lambda$ from $\Lambda_0$ to $0$. In practice the parameter $\Lambda$ is chosen as an infrared regularization which effectively removes low-energy (or low-momentum) modes, determined by the cutoff $\Lambda$, from the functional integration. In this case, the initial value $\Lambda_0$ is the ultraviolet cutoff of the action $S[\boldsymbol{\psi},\boldsymbol{\phi}]$. For graphene, this ultraviolet cutoff is the momentum or energy at which the linear dispersion in Eq.~\eqref{eq:FreeHamiltonian} breaks down.
\subsection{Infrared regularization}
\label{sec:InfraredRegularization}
We implement the idea of an infrared regularization by modifying the quadratic terms in the Fermi- and Bose-sectors of the contour-time action via additive regulator functions $\hat{R}_{\textrm{f}, \Lambda}, R_{\textrm{b}, \Lambda}$~\cite{Wetterich1991, BergesTetradisWetterich2002}
\begin{subequations}
\begin{align}
&S_{\textrm{f}}[\psi] \rightarrow S_{\textrm{f}, \Lambda}[\psi] = S_{\textrm{f}}[\psi] + \Psi^{\dagger} \hat{R}_{\textrm{f}, \Lambda} \Psi \,, \\
&S_{\textrm{b}}[\phi] \rightarrow S_{\textrm{b}, \Lambda}[\phi] = S_{\textrm{b}}[\phi] + \frac{1}{2} \phi^{\intercal} R_{\textrm{b}, \Lambda} \phi \,.
\label{eq:ReplacementQuadraticPartInfraredRegularization}
\end{align}
\end{subequations}
It is also possible to regularize only one of the two sectors, by setting either $\hat{R}_{\textrm{f}, \Lambda}$ or $R_{\textrm{b}, \Lambda}$ to zero. The regulators have to be analytic functions of $\Lambda$. For $\Lambda \rightarrow \Lambda_0$ they have to diverge, such that all infrared modes occuring in the functional integral are effectively frozen out, while for $\Lambda \rightarrow 0$ they have to vanish.\cite{BergesTetradisWetterich2002, MetznerHonercampEtal2012, KopietzBook} In this way the partition function \eqref{eq:FermiBoseGeneratingFunctional} becomes a cutoff dependent quantity, $Z[\eta, J] \rightarrow Z_{\Lambda}[\eta, J]$, where only the modes above $\Lambda$ contribute to the functional integral. In the limit $\Lambda \rightarrow 0$ it reduces to the original partition function of the previous section, see Eq.~\eqref{eq:KeldyshFermiBoseGeneratingFunctional}.



After mapping the contour-time regulator terms to a real-time repesentation and performing the Keldysh rotation as explained in Sec.~\ref{sec:RealTimeRepresentation}, the cutoff dependent quadratic parts of the action become
\begin{subequations} \label{eq:KeldyshRegulator}
\begin{align}
S_{\textrm{f}, \Lambda}[\boldsymbol{\psi}] &= S_{\textrm{f}}[\boldsymbol{\psi}] + \boldsymbol{\Psi^{\dagger}} \boldsymbol{\hat{R}}_{\textrm{f}, \Lambda} \boldsymbol{\Psi} \,,
\label{eq:KeldyshRegulatorFermions} \\
S_{\textrm{b}, \Lambda}[\boldsymbol{\phi}] &= S_{\textrm{b}}[\boldsymbol{\phi}] + \boldsymbol{\phi}^{\intercal} \boldsymbol{R}_{\textrm{b}, \Lambda} \boldsymbol{\phi} \,.
\label{eq:KeldyshRegulatorBoson}
\end{align}
\end{subequations}
Note the absence of the factor 1/2 in front of the bosonic regulator term, which is due to our choice of normalization for the bosonic rotation~\eqref{eq:DefinitionClassicalQuantumFieldsBosons}. In principle, the most general choice for the contour-time regulators results in the following $2 \times 2$ matrix structure for the real-time regulators
\begin{subequations}
\begin{align}
\boldsymbol{\hat{R}}_{\textrm{f}, \Lambda}(x, y) &= \begin{pmatrix} \hat{R}_{\textrm{f}, \Lambda}^Z(x, y) & \hat{R}_{\textrm{f}, \Lambda}^A(x, y) \\ \hat{R}_{\textrm{f}, \Lambda}^R(x, y) & \hat{R}_{\textrm{f}, \Lambda}^K(x, y) \end{pmatrix} \,,
\label{eq:KeldyshStructureFermionicRegulators} \\
\boldsymbol{R}_{\textrm{b}, \Lambda}(x, y) &= \begin{pmatrix} R_{\textrm{b}, \Lambda}^Z(x, y) & R_{\textrm{b}, \Lambda}^A(x, y) \\ R_{\textrm{b}, \Lambda}^R(x, y) & R_{\textrm{b}, \Lambda}^K(x, y) \end{pmatrix} \,.
\label{eq:KeldyshStructureBosonicRegulators}
\end{align}
\end{subequations}

Although it is not strictly necessary if the evolution from $\Lambda = \Lambda_0$ to $\Lambda = 0$ could be tracked exactly, for the correct implementation of approximate evolution schemes it is important that the regulators are chosen in such a way that they respect the symmetries and the causality structure of the theory. In particular, in order to ensure that the partition function is normalized to unity at any scale, and hence retain the algebraic identities among the correlation functions, \textit{cf.} Eq.~\eqref{eq:GFConstraint}, we choose the regulators such that the ``anomalous'' components $\hat{R}_{\textrm{f}, \Lambda}^Z, R_{\textrm{b}, \Lambda}^Z$ vanish. The remaining components are constructed in such a way that they are compatible with the symmetry and causality structure of the bare inverse propagators, see Eqs.~\eqref{eq:SymmetriesInverseFreePropagatorsFermions} and \eqref{eq:SymmetriesInverseFreePropagatorsBoson2}. This choice of the regulator functions ensures that the partition function has the correct causality structure at any value of the cutoff $\Lambda$, independent of eventual approximations made when solving the evolution equations.

In addition to the $\Lambda$-dependencee of the action introduced via Eqs.~\eqref{eq:KeldyshRegulator} we allow the counterterm to be explicitly cutoff dependent, setting 
\begin{equation}
\tilde{n} \rightarrow \tilde{n}_{\Lambda}
\label{eq:CutoffDependentCounterTerm}
\end{equation} 
The counterterm $\tilde{n}_{\Lambda}$ describes a flowing background charge density, which has to be tuned to remove potentially divergent contributions from the Coulomb interaction at finite charge carrier density.

\subsection{Connected functional and effective action}
\label{sec:ConnectedFunctionalAndEffectiveAction}
The evolution equation will not be derived for the partition function $Z_{\Lambda}[\boldsymbol{\eta},\boldsymbol{J}]$, but rather for the effective action $\Gamma_{\Lambda}[\boldsymbol{\psi},\boldsymbol{\phi}]$, which is essentially the Legendre transformation of the cutoff dependent connected functional~\cite{NegeleOrlandBook, ChouSuYu1985}
\begin{equation}
W_{\Lambda}[\boldsymbol{\eta}, \boldsymbol{J}] = - i \textrm{ln} Z_{\Lambda}[\boldsymbol{\eta}, \boldsymbol{J}] \,,
\label{eq:ConnectedFunctional}
\end{equation}
being a generating functional for connected correlation functions. Differentiation with respect to the sources yield the expectation values of the fields $\boldsymbol{\Psi}(x)$ and $\boldsymbol{\phi}(x)$,
\begin{align}
\frac{\delta W_{\Lambda}}{\delta \boldsymbol{\eta^{\dagger}}(x)} &= + \tau_1 \big\langle \boldsymbol{\Psi}(x) \big\rangle \,, \quad
\frac{\delta W_{\Lambda}}{\delta \boldsymbol{\eta}(x)} = - \big\langle \boldsymbol{\Psi^{\dagger}}(x) \big\rangle \tau_1 \,, 
\label{eq:DefinitionFieldExpectationValuesFermions} \\
\frac{\delta W_{\Lambda}}{\delta \boldsymbol{J}(x)} &= \big\langle \boldsymbol{\phi}^{\intercal}(x) \big\rangle \tau_1 \,.
\label{eq:DefinitionFieldExpectationValueBoson}
\end{align}
%
%
These expectation values, being complicated nonlinear functionals of the sources $\boldsymbol{\eta}$ and $\boldsymbol{J}$, define ``macroscopic'' fields which inherit a $\Lambda$-dependence from the regulators (and the counterterm). A macroscopic Fermi field can only exist when the sources are finite, otherwise it is strictly zero. The classical component of the macroscopic bosonic field $\langle \phi_c(x) \rangle$, on the other hand, can very well acquire a finite value in the absence of source terms.\cite{ChouSuYu1985, KlossKopietz2011, BergesMesterhazy2012, BergesHoffmeister2009} Such a macroscopic field expectation value may signal a spontaneous symmetry breaking, but in the theory we consider here this is not the case. The bosonic field $\boldsymbol{\phi}(x)$ is conjugate to the particle density $n(x)$ and as such it reflects, \textit{e.g.}, a local deviation away from charge neutrality driven by an external potential $\varphi(x)$. In the following we omit the brackets to denote the average of a single field, for brevity. Since we are always working with averages of fields, there can be no confusion.

The second derivatives of $W_{\Lambda}$ define the connected two-point correlators
\begin{align}
\frac{\delta^2 W_{\Lambda}}{\delta \boldsymbol{\eta^{\dagger}}(x) \delta \boldsymbol{\eta}(y)} &= - i \tau_1 \big\langle \boldsymbol{\Psi}(x) \boldsymbol{\Psi^{\dagger}}(y) \big\rangle_{\textrm{c}} \, \tau_1 \,, 
\label{eq:ConnectedCorrelatorsDerivativeFermions} \\
\frac{\delta^2 W_{\Lambda}}{\delta \boldsymbol{J}^{\intercal}(x) \delta \boldsymbol{J}(y)} &= + i \tau_1 \big\langle \boldsymbol{\phi}(x) \boldsymbol{\phi^{\intercal}}(y) \big\rangle_{\textrm{c}} \, \tau_1 \,,
\label{eq:ConnectedCorrelatorsDerivativeBoson}
\end{align}
where we introduced the connected average $\langle AB \rangle_{\textrm{c}} \equiv \langle AB \rangle - \langle A \rangle \langle B \rangle$. Explicitly displaying the $2 \times 2$ Keldysh structure, we have
\begin{align}
\langle \boldsymbol{\Psi}(x) \boldsymbol{\Psi^{\dagger}}(y) \rangle_{\textrm{c}} &\equiv i \boldsymbol{\hat{G}}_{\Lambda}(x, y| \boldsymbol{\eta}, \boldsymbol{J})
\label{eq:ConnectedCorrelatorsFermions} \\
&= i \begin{pmatrix} \hat{G}_{\Lambda}^K(x, y| \boldsymbol{\eta}, \boldsymbol{J}) & \hat{G}_{\Lambda}^R(x, y| \boldsymbol{\eta}, \boldsymbol{J}) \\ \hat{G}_{\Lambda}^A(x, y| \boldsymbol{\eta}, \boldsymbol{J}) & \hat{G}_{\Lambda}^Z(x, y| \boldsymbol{\eta}, \boldsymbol{J}) \end{pmatrix} \,, \nonumber \\[0.25cm]
\langle \boldsymbol{\phi}(x) \boldsymbol{\phi}^{\intercal}(y) \rangle_{\textrm{c}} &\equiv i \boldsymbol{D}_{\Lambda}(x, y| \boldsymbol{\eta}, \boldsymbol{J}) 
\label{eq:ConnectedCorrelatorsBoson} \\
&= i \begin{pmatrix} D_{\Lambda}^K(x, y| \boldsymbol{\eta}, \boldsymbol{J}) & D_{\Lambda}^R(x, y| \boldsymbol{\eta}, \boldsymbol{J}) \\ D_{\Lambda}^A(x, y| \boldsymbol{\eta}, \boldsymbol{J}) & D_{\Lambda}^Z(x, y| \boldsymbol{\eta}, \boldsymbol{J}) \end{pmatrix} \,. \nonumber
\end{align}
%
%
The above propagators are source- and cutoff-dependent functionals, which do not obey the usual triangular structure.
In particular the anomalous statistical propagators $\hat{G}_{\Lambda}^Z$ and $D_{\Lambda}^Z$ are nonvanishing as long as the source terms are finite. By construction of the regulators, the familiar triangular structure together with the symmetry and causality relations arise once the single-particle sources are set to zero.
All the other higher order connected correlation functions can be obtained by further differentiation as in the equilibrium Matsubara theory.\cite{NegeleOrlandBook}

The central object in the functional renormalization group is the effective action $\Gamma_{\Lambda}[\boldsymbol{\psi}, \boldsymbol{\phi}]$. It is the generating functional for one-particle irreducible vertex functions, and defined as the Legendre transform of the connected functional~$W_{\Lambda}$
\begin{align}
\Gamma_{\Lambda}[\boldsymbol{\psi}, \boldsymbol{\phi}] = &W_{\Lambda}[\boldsymbol{\eta}_{\Lambda}, \boldsymbol{J}_{\Lambda}] - \boldsymbol{\eta^{\dagger}}_{\Lambda} \tau_1 \boldsymbol{\Psi} - \boldsymbol{\Psi^{\dagger}} \tau_1 \boldsymbol{\eta}_{\Lambda} - \boldsymbol{\phi}^{\intercal} \tau_1 \boldsymbol{J}_{\Lambda} \nonumber \\
&- \vec{\boldsymbol{\Psi}}^{\dagger} \boldsymbol{\hat{R}}_{\textrm{f}, \Lambda} \vec{\boldsymbol{\Psi}} - \boldsymbol{\phi}^{\intercal} \boldsymbol{R}_{\textrm{b}, \Lambda} \boldsymbol{\phi} \,.
\label{eq:LegendreTransform}
\end{align}
In the Legendre transform the single-particle sources must be understood as $\Lambda$-dependent functionals of the field expectation values, obtained by inversion of the defining relations Eqs.~\eqref{eq:DefinitionFieldExpectationValuesFermions} and \eqref{eq:DefinitionFieldExpectationValueBoson}. The Legendre transform is modified in such a way that the cutoff terms are subtracted on the right hand side. This ensures that the flowing action does not contain the cutoff terms at any scale, but spoils the convexity of an ordinary Legendre transform.

The properties and physical interpretation of this functional, mainly in the context of its equilibrium counterpart, have been discussed at length in the literature.\cite{BergesTetradisWetterich2002, MetznerHonercampEtal2012, KopietzBook} Most importantly the flowing action has the nice property that it interpolates smoothly between the microscopic laws of physics, parametrized by an action $\Gamma_{\Lambda_0}$, and the full effective action $\Gamma_{\Lambda = 0}$, where all thermal and quantum fluctuations are taken into account. In many cases the microscopic laws are simply governed by the bare action of the system $\Gamma_{\Lambda_0} = S$. This latter statement, however, depends on the actual cutoff scheme. In certain situations it is preferable to devise a cutoff scheme where the initial effective action does not coincide with the bare action, and hence the initial conditions of the flow are nontrivial.\cite{JakobsPletyukovSchoeller2010, JakobsMedenSchoeller2007, KopietzBook, SchuetzKopietz2006, SharmaKopietz2016} We will come back to this issue at the end of the next subsection. 

Taking the first functional derivative of Eq.~\eqref{eq:LegendreTransform} with respect to the fields one finds that the effective action satisfies the ``equations of motion''
\begin{align}
\frac{\delta \Gamma_{\Lambda}}{\delta \boldsymbol{\Psi^{\dagger}}(x)} &= - \tau_1 \boldsymbol{\eta}_{\Lambda}(x) - \int_y \boldsymbol{\hat{R}}_{\textrm{f}, \Lambda}(x, y) \boldsymbol{\Psi}(y) \,, 
\label{eq:EquationsOfMotionGamma1} \\
\frac{\delta \Gamma_{\Lambda}}{\delta \boldsymbol{\Psi}(x)} &= + \boldsymbol{\eta^{\dagger}}_{\Lambda}(x) \tau_1 + \int_y \boldsymbol{\Psi^{\dagger}}(y) \boldsymbol{\hat{R}}_{\textrm{f}, \Lambda}(y, x) \,, 
\label{eq:EquationsOfMotionGamma2} \\
\frac{\delta \Gamma_{\Lambda}}{\delta \boldsymbol{\phi}^{\intercal}(x)} &= - \tau_1 \boldsymbol{J}_{\Lambda}(x) - 2 \int_y \boldsymbol{R}_{\textrm{b}, \Lambda}(x, y) \boldsymbol{\phi}(y) \,.
\label{eq:EquationsOfMotionGamma3}
\end{align}
The second functional derivatives of the connected functional $W_{\Lambda}[\boldsymbol{\eta}, \boldsymbol{J}]$ and the second functional derivatives of the effective action $\Gamma_{\Lambda}[\boldsymbol{\psi}, \boldsymbol{\phi}]$ are subject to an inversion relation,\cite{NegeleOrlandBook, KopietzBook, MetznerHonercampEtal2012} which can be written in the compact form
\begin{equation}
- \left( \boldsymbol{\hat{\Gamma}}_{\Lambda}^{(2)} + \boldsymbol{\hat{\mathcal{R}}}_{\Lambda} \right) \boldsymbol{\tau_1} \boldsymbol{\hat{W}}_{\Lambda}^{(2)} \boldsymbol{\tau_1} = \hat{\mathbb{1}} \,.
\label{eq:GeneralizedDysonEquation}
\end{equation}
Here we have defined the matrices
\begin{align}
\boldsymbol{\hat{\mathcal{R}}}_{\Lambda} &\equiv \textrm{diag} \left( - \boldsymbol{\hat{R}}_{\textrm{f}, \Lambda}, \boldsymbol{\hat{R}}_{\textrm{f}, \Lambda}^{\intercal}, 2 \boldsymbol{R}_{\textrm{b}, \Lambda} \right) \,, \\
\boldsymbol{\tau_1} &\equiv \textrm{diag}(\tau_1, \tau_1, \tau_1) \,,
\label{eq:DefinitionRegulatorMatrixTauMatrix}
\end{align}
and the Hesse matrices $\boldsymbol{\hat{W}}_{\Lambda}^{(2)}$ and $\boldsymbol{\hat{\Gamma}}_{\Lambda}^{(2)}$ of second functional derivatives
\begin{align}
\boldsymbol{\hat{W}}_{\Lambda}^{(2)} &=
\begin{pmatrix} 
\delta_{\boldsymbol{\eta^{\dagger}}} \delta_{\boldsymbol{\eta}} & - \delta_{\boldsymbol{\eta^{\dagger}}} \delta_{\boldsymbol{\eta^{\dagger}}}^{\intercal} & - \delta_{\boldsymbol{\eta^{\dagger}}} \delta_{\boldsymbol{J}} \\[1mm]
- \delta_{\boldsymbol{\eta}}^{\intercal} \delta_{\boldsymbol{\eta}} & \delta_{\boldsymbol{\eta}}^{\intercal} \delta_{\vec{\boldsymbol{\eta}}^{\dagger}}^{\intercal} & \delta_{\vec{\boldsymbol{\eta}}}^{\intercal} \delta_{\boldsymbol{J}} \\[1mm]
- \delta_{\boldsymbol{J}}^{\intercal} \delta_{\boldsymbol{\eta}} & \delta_{\boldsymbol{J}}^{\intercal} \delta_{\boldsymbol{\eta^{\dagger}}}^{\intercal} & \delta_{\boldsymbol{J}}^{\intercal} \delta_{\boldsymbol{J}}
\end{pmatrix}
W_{\Lambda} \,,
\label{eq:SecondFunctionalDerivativeMatrixW} \\
\boldsymbol{\hat{\Gamma}}_{\Lambda}^{(2)} &=
\begin{pmatrix}
\delta_{\boldsymbol{\Psi^{\dagger}}} \delta_{\boldsymbol{\Psi}} & \delta_{\boldsymbol{\Psi^{\dagger}}} \delta_{\boldsymbol{\Psi^{\dagger}}}^{\intercal} & \delta_{\boldsymbol{\Psi^{\dagger}}} \delta_{\boldsymbol{\phi}} \\[1mm]
\delta_{\boldsymbol{\Psi}}^{\intercal} \delta_{\boldsymbol{\Psi}} & \delta_{\boldsymbol{\Psi}}^{\intercal} \delta_{\vec{\boldsymbol{\Psi}}^{\dagger}}^{\intercal} & \delta_{\vec{\boldsymbol{\Psi}}}^{\intercal} \delta_{\boldsymbol{\phi}} \\[1mm]
\delta_{\boldsymbol{\phi}}^{\intercal} \delta_{\boldsymbol{\Psi}} & \delta_{\boldsymbol{\phi}}^{\intercal} \delta_{\boldsymbol{\Psi^{\dagger}}}^{\intercal} & \delta_{\boldsymbol{\phi}}^{\intercal} \delta_{\boldsymbol{\phi}}
\end{pmatrix}
\Gamma_{\Lambda} \,.
\label{eq:SecondFunctionalDerivativeMatrixGamma}
\end{align}
The inversion relation~\eqref{eq:GeneralizedDysonEquation} generalizes the standard Dyson equations for single-particle propagators to the case of source-dependent functional propagators, Eqs.~\eqref{eq:ConnectedCorrelatorsFermions}, \eqref{eq:ConnectedCorrelatorsBoson}. If the sources are finite, Eq.~\eqref{eq:GeneralizedDysonEquation} also includes mixed Fermi-Bose correlators, which disappear upon setting the sources to zero. Applying further functional derivatives to this equation yields a tree expansion of a connected $n$-particle correlation function in terms $m$-particle vertex functions ($m \leq n$) and full propagators; see Refs.~\onlinecite{NegeleOrlandBook, KopietzBook, MetznerHonercampEtal2012}.

\subsection{Dyson and quantum kinetic equations in the functional renormalization group}
\label{sec:DysonAndQuantumKineticEquationsInTheFunctionalRenormalizationGroup}
Evaluating the generalized Dyson equation at vanishing sources we obtain the scale dependent nonequilibrium Dyson equations for Fermions and Bosons
\begin{align}
\left( \boldsymbol{\hat{G}}_0^{-1} - \boldsymbol{\hat{\Sigma}}_{\Lambda} + \boldsymbol{\hat{R}}_{\textrm{f}, \Lambda} \right) \boldsymbol{\hat{G}}_{\Lambda} = \hat{\mathbb{1}} \,,
\label{eq:DysonEquationFermions} \\
2 \Big( \boldsymbol{V}^{-1} + \boldsymbol{\Pi}_{\Lambda} + \boldsymbol{R}_{\textrm{b}, \Lambda} \Big) \boldsymbol{D}_{\Lambda} = \mathbb{1} \,,
\label{eq:DysonEquationBoson}
\end{align}
where we employed the definition of the unregularized inverse full propagators
\begin{align}
\frac{\delta^2 \Gamma_{\Lambda}}{\delta \boldsymbol{\Psi^{\dagger}}(x) \delta \boldsymbol{\Psi}(y)} \bigg|_{\phi_c = \bar{\phi}_c} &= - \Big(\boldsymbol{\hat{G}}_0^{-1} - \boldsymbol{\hat{\Sigma}}_{\Lambda} \Big)(x, y) \,, 
\label{eq:SelfenergyDefinitionFermions} \\
\frac{\delta^2 \Gamma_{\Lambda}}{\delta \boldsymbol{\phi^{\intercal}}(x) \delta \boldsymbol{\phi}(y)}\bigg|_{\phi_c = \bar{\phi}_c} &= 2 \Big(\boldsymbol{V}^{-1} + \boldsymbol{\Pi}_{\Lambda} \Big)(x, y) \,.
\label{eq:SelfenergyDefinitionBoson}
\end{align}
Here Eqs.~\eqref{eq:SelfenergyDefinitionFermions} and \eqref{eq:SelfenergyDefinitionBoson} define the (fermionic) self-energy $\boldsymbol{\hat{\Sigma}}_{\Lambda}$ and the (bosonic) polarization function $\boldsymbol{\Pi}_{\Lambda}$, respectively. The latter is also known as bosonic self-energy, which will be used synonymously in the remainder of this work.\cite{Note4}
%
%
By construction of the regulators, the fermionic and bosonic self-energies have the same trigonal structure in Keldysh space as the inverse free propagators and the regulators
\begin{align}
\boldsymbol{\hat{\Sigma}}_{\Lambda} = \begin{pmatrix} 0 & \hat{\Sigma}_{\Lambda}^A \\ \hat{\Sigma}_{\Lambda}^R & \hat{\Sigma}_{\Lambda}^K  \end{pmatrix} \,, \quad \boldsymbol{\Pi}_{\Lambda} = \begin{pmatrix} 0 & \Pi_{\Lambda}^A \\ \Pi_{\Lambda}^R & \Pi_{\Lambda}^K  \end{pmatrix} \,.
\label{eq:FermionicBosonicSelfenergiesKeldyshStructure}
\end{align}
Besides, they inherit their causality and symmetry relations, see Eqs.~\eqref{eq:SymmetriesInverseFreePropagatorsFermions} and \eqref{eq:SymmetriesInverseFreePropagatorsBoson2}.
%
%

The diagonal components of Eqs.~\eqref{eq:DysonEquationFermions} and \eqref{eq:DysonEquationBoson} contain the respective retarded and advanced Dyson equations
\begin{align}
\left( \big( \hat{G}_0^{R/A} \big)^{-1} - \hat{\Sigma}_{\Lambda}^{R/A} + \hat{R}_{\textrm{f}, \Lambda}^{R/A} \right) \hat{G}_{\Lambda}^{R/A} &= \hat{\mathbb{1}} \,,
\label{eq:DiagonalComponentsDysonEquationFermions} \\
2 \Big( V^{-1} + \Pi_{\Lambda}^{R/A} + R_{\textrm{b}, \Lambda}^{R/A} \Big) D^{R/A} &= \mathbb{1} \,,
\label{eq:DiagonalComponentsDysonEquationBoson}
\end{align}
whereas their off-diagonal yield the Keldysh Green functions
\begin{align}
\hat{G}_{\Lambda}^K &= - \hat{G}_{\Lambda}^R \left( - \hat{\Sigma}_{\Lambda}^K + \hat{R}_{\textrm{f}, \Lambda}^K \right) \hat{G}_{\Lambda}^A \,,
\label{eq:InteractingKeldyshPropagatorFermions} \\
D_{\Lambda}^K &= - 2 D_{\Lambda}^R \Big( \Pi_{\Lambda}^K + R_{\textrm{b}, \Lambda}^K \Big) D_{\Lambda}^A \,.
\label{eq:InteractingKeldyshPropagatorBoson}
\end{align}
These relations are a straightforward generalization of Eqs.~\eqref{eq:KeldyshComponentInverseFreePropagatorFermions} and \eqref{eq:KeldyshComponentInverseFreePropagatorBoson} to the interacting case and the presence of infrared regulators.

Continuing the parallels with the noninteracting case, the flowing full Keldysh propagators $\hat{G}_{\Lambda}^K$ and $D_{\Lambda}^K$ can be paramterized in terms of cutoff dependent hermitian matrices $\hat{\mathcal{F}}_{\Lambda}$ and $\mathcal{B}_{\Lambda}$, respectively,
\begin{align}
\hat{G}_{\Lambda}^K = \hat{G}_{\Lambda}^R \hat{\mathcal{F}}_{\Lambda} - \hat{\mathcal{F}}_{\Lambda} \hat{G}_{\Lambda}^A \,, \\
D_{\Lambda}^K = D_{\Lambda}^R \mathcal{B}_{\Lambda} - \mathcal{B}_{\Lambda} D_{\Lambda}^A \,.
\label{eq:ParametrizationFlowingKeldyshPropagators}
\end{align}
This paramterization can be used to derive an equation of motion for each of the distribution functions $\hat{\mathcal{F}}_{\Lambda}$ and $\mathcal{B}_{\Lambda}$. Such equations of motion are known as the quantum kinetic equations.\cite{KamenevBook} To this end we insert the above parametrization into Eq.~\eqref{eq:InteractingKeldyshPropagatorFermions}, respectively Eq.~\eqref{eq:InteractingKeldyshPropagatorBoson}. Applying the retarded inverse full propagator from the left and the advanced one from the right, we obtain the two kinetic equations
\begin{widetext}
\begin{align}
&\Big[ \hat{\mathcal{F}}_{\Lambda}, \hat{G}_0^{-1} \Big] + \hat{R}_{\textrm{f}, \Lambda}^K \,- \Big( \hat{R}_{\textrm{f}, \Lambda}^R \hat{\mathcal{F}}_{\Lambda} - \hat{\mathcal{F}}_{\Lambda} \hat{R}_{\textrm{f}, \Lambda}^A \Big) \, = \hat{\Sigma}_{\Lambda}^K - \Big( \hat{\Sigma}_{\Lambda}^R \hat{\mathcal{F}}_{\Lambda} - \hat{\mathcal{F}}_{\Lambda} \hat{\Sigma}_{\Lambda}^A \Big) \,,
\label{eq:QuantumKineticEquationFermions} \\
&\Big[ \mathcal{B}_{\Lambda}, V^{-1} \Big] + R_{\textrm{b}, \Lambda}^K - \Big( R_{\textrm{b}, \Lambda}^R \mathcal{B}_{\Lambda} - \mathcal{B}_{\Lambda} R_{\textrm{b}, \Lambda}^A \Big) = -\Pi_{\Lambda}^K + \Big( \Pi_{\Lambda}^R \mathcal{B}_{\Lambda} - \mathcal{B}_{\Lambda} \Pi_{\Lambda}^A \Big) \,,
\label{eq:QuantumKineticEquationBoson}
\end{align}
\end{widetext}
where $[\cdot, \cdot]$ denotes the commutator. The left hand side of these equations is the kinetic term, while their right hand side is known as the collision integral. Note that the commutator for the bosonic distribution function $\mathcal{B}_{\Lambda}$ does not involve any time derivatives: The dynamics of $\mathcal{B}_{\Lambda}$ is entirely driven by the bosonic collision integral, and thus induced by the dynamics of the fermions. In a general nonequilibrium situation the kinetic terms do not vanish and, hence, the Keldysh self-energies do not admit the same decomposition as the Keldysh propagators, leading to a finite collision integral. 

We want to stress that the $\Lambda$-dependence of the distribution functions, since it is a parametric one, poses a serious complication. The kinetic equations have to be solved at each scale, together with the flow equations for the various self-energies and higher order vertex functions, selfconsistently. The latter set of flow equations will be derived in the next subsections. Therefore further approximations are inevitable, if one hopes to obtain numerical solutions for a specific nonequilibrium problem. For example, if the external fields are taken to be slowly varying functions of time and/or space, one could use a Wigner transformation and perform a gradient expansion to some low order.\cite{BergesBorsanyi2006, KamenevBook} Often this approximation is combined with the so-called quasiparticle approximation, which reduces the phase space of the distribution functions and eventually leads to the Boltzmann transport equation. An important technical simplification is achieved by the class of cutoff schemes where the Keldysh regulators are parametrized in the same way as the Keldysh propagators 
\begin{align}
\hat{R}_{\textrm{f}, \Lambda}^K &= \hat{R}_{\textrm{f}, \Lambda}^R \hat{\mathcal{F}}_{\Lambda} - \hat{\mathcal{F}}_{\Lambda} \hat{R}_{\textrm{f}, \Lambda}^A \,, 
\label{eq:ParametrizationKeldyshRegulatorsFermions} \\
R_{\textrm{b}, \Lambda}^K &= R_{\textrm{b}, \Lambda}^R \mathcal{B}_{\Lambda} - \mathcal{B}_{\Lambda} R_{\textrm{b}, \Lambda}^A \,.
\label{eq:ParametrizationKeldyshRegulatorsBoson}
\end{align}
As a consequence the regulators on the left hand side of the above kinetic equations drop out and we are left with the kinetic equations in their standard form, as if no regulators were present, see Ref.~\onlinecite{KamenevBook}. Especially in the treatment of equilibrium problems this fact has a great advantage. Namely, it is possible to solve the kinetic equations at all scales simultaneously with the well-known equilibrium distribution functions. In this way the results of the Matsubara formalism are reproduced directly in real time, avoiding the necessity of cumbersome analytic continuations. We will come back to the equilibrium problem in the final section of this article, Sec.~\ref{sec:ThermalEquilibrium}. The drawback of these schemes, however, is that the initial conditions of the flow equations, become nontrivial as pointed out in the Refs.~\onlinecite{BergesMesterhazy2012, BergesHoffmeister2009, JakobsMedenSchoeller2007, JakobsPletyukovSchoeller2010}, meaning that $\Gamma_{\Lambda}$ in the limit $\Lambda \rightarrow \Lambda_0$ does not coincide with the bare action $S$. On the other hand this is a rather small price to pay.

\subsection{Exact flow equation}
\label{sec:ExactFlowEquation}
The implementation of the infrared regulators described above enables us to derive an exact evolution equation for the effective action $\Gamma_{\Lambda}$, which describes its flow in the infinite dimensional space of all possible actions as a function of the flowing cutoff $\Lambda$. The flow equation for $\Gamma_{\Lambda}$ follows upon taking the $\Lambda$-derivative of the defining relation, Eq.~\eqref{eq:LegendreTransform}, at a fixed field configuration. To this end recall that the connected functional $W_{\Lambda}[\boldsymbol{\eta}_{\Lambda}, \boldsymbol{J}_{\Lambda}]$ therein has an explicit and an implicit $\Lambda$-dependence. The flow of the sources $\boldsymbol{\eta}_{\Lambda}$ and $\boldsymbol{J}_{\Lambda}$, viewed as functionals of the fields $\boldsymbol{\Psi}$ and $\boldsymbol{\phi}$, does not contribute to the flow of $\Gamma_{\Lambda}$ as the respective terms cancel each other.\cite{BergesTetradisWetterich2002, MetznerHonercampEtal2012} We thus find
\begin{equation}
\partial_{\Lambda} \Gamma_{\Lambda} = \partial_{\Lambda} W_{\Lambda} - \vec{\boldsymbol{\Psi}}^{\dagger} \partial_{\Lambda} \boldsymbol{\hat{R}}_{\textrm{f}, \Lambda} \vec{\boldsymbol{\Psi}} - \boldsymbol{\phi}^{\intercal} \partial_{\Lambda} \boldsymbol{R}_{\textrm{b}, \Lambda} \boldsymbol{\phi} \,,
\label{eq:DerivativeOfLegendreTransform}
\end{equation}
where the scale-derivative of the first term on the right hand side, $\partial_{\Lambda} W_{\Lambda}$, has to be performed for fixed source fields $\boldsymbol{\eta}$ and $\boldsymbol{J}$. It obeys an exact flow equation as well, which is readily derived from the definition \eqref{eq:ConnectedFunctional}
\begin{widetext}
\begin{align}
\partial_{\Lambda} W_{\Lambda} &= \left \langle \boldsymbol{\Psi}^{\dagger} \partial_{\Lambda} \boldsymbol{\hat{R}}_{\textrm{f}, \Lambda} \boldsymbol{\Psi} \right \rangle + \Big \langle \boldsymbol{\phi}^{\intercal} \partial_{\Lambda} \boldsymbol{R}_{\textrm{b}, \Lambda} \boldsymbol{\phi} \Big \rangle + 2 \boldsymbol{\phi}^{\intercal} \tau_1 \partial_{\Lambda} \boldsymbol{\tilde{n}}_{\Lambda} \nonumber \\
&= - \textrm{Tr} \left( (\partial_{\Lambda} \boldsymbol{\hat{R}}_{\textrm{f}, \Lambda}) \Big( \big\langle \boldsymbol{\Psi} \boldsymbol{\Psi}^{\dagger} \big\rangle_{\textrm{c}} + \boldsymbol{\Psi} \boldsymbol{\Psi}^{\dagger} \Big) \right) + \textrm{Tr} \Big( (\partial_{\Lambda} \boldsymbol{R}_{\textrm{b}, \Lambda}) \left( \big\langle \boldsymbol{\phi} \boldsymbol{\phi}^{\intercal} \big\rangle_{\textrm{c}} + \boldsymbol{\phi} \boldsymbol{\phi}^{\intercal} \Big) \right) +2 \boldsymbol{\phi}^{\intercal} \tau_1 \partial_{\Lambda} \boldsymbol{\tilde{n}}_{\Lambda} \,.
\label{eq:ExactFlowEquationForW}
\end{align}
\end{widetext}
Here the trace $\textrm{Tr}$ encompasses an integration over position and time, as well as a summation over the Keldysh components $c$ and $q$ and, for fermions, a summation over the spin, valley and sublattice indices. Note the occurence of the flowing counterterm $\boldsymbol{\tilde{n}}_{\Lambda}$ on the right hand side, and recall that it possesses a classical component only. Upon insertion of Eq.~\eqref{eq:ExactFlowEquationForW} into Eq.~\eqref{eq:DerivativeOfLegendreTransform} the additional regulator terms cancel, such that the flow equation contains connected functional propagators and the counterterm only. Making use of Eqs.~\eqref{eq:ConnectedCorrelatorsDerivativeFermions}, \eqref{eq:ConnectedCorrelatorsDerivativeBoson} and~\eqref{eq:SecondFunctionalDerivativeMatrixW}, we can write our intermediate result in the compact form
\begin{equation}
\partial_{\Lambda} \Gamma_{\Lambda} = - \frac{i}{2} \textrm{STr} \left( (\partial_{\Lambda} \boldsymbol{\hat{\mathcal{R}}}_{\Lambda}) \boldsymbol{\tau_1} \boldsymbol{\hat{W}}_{\Lambda}^{(2)} \boldsymbol{\tau_1} \right) + 2 \boldsymbol{\phi}^{\intercal} \tau_1 \partial_{\Lambda} \boldsymbol{\tilde{n}}_{\Lambda} \,.
\label{eq:ExactFlowEquationForGammaIntermediate}
\end{equation}
We recognize here the well-known one-loop structure of the flow equation with the cutoff-insertion $\partial_{\Lambda} \boldsymbol{\hat{\mathcal{R}}}_{\Lambda}$. The usual minus sign for a closed fermion loop has been absorbed into the definition
\begin{equation}
\textrm{STr}( \cdots ) \equiv \textrm{Tr} \left( \begin{pmatrix} -1 & 0 & 0 \\ 0 & -1 & 0 \\ 0 & 0 & 1 \end{pmatrix} \cdots \right) \,.
\label{eq:DefinitionSTrace}
\end{equation}
In order to close Eq.~\eqref{eq:ExactFlowEquationForGammaIntermediate} we make use of the generalized Dyson equation \eqref{eq:GeneralizedDysonEquation} and write
\begin{align}
\partial_{\Lambda} \Gamma_{\Lambda} = \,&\frac{i}{2} \textrm{STr} \left( (\partial_{\Lambda} \boldsymbol{\hat{\mathcal{R}}}_{\Lambda}) \left( \boldsymbol{\hat{\Gamma}}_{\Lambda}^{(2)} + \boldsymbol{\hat{\mathcal{R}}}_{\Lambda} \right)^{-1} \right) \nonumber \\ 
&+ 2 \boldsymbol{\phi}^{\intercal} \tau_1 \partial_{\Lambda} \boldsymbol{\tilde{n}}_{\Lambda} \nonumber \\
= \,&\frac{i}{2} \partial \!\!\!/_{\Lambda} \textrm{STr} \, \textrm{ln} \left( \boldsymbol{\hat{\Gamma}}_{\Lambda}^{(2)} + \boldsymbol{\hat{\mathcal{R}}}_{\Lambda} \right) + 2 \boldsymbol{\phi}^{\intercal} \tau_1 \partial_{\Lambda} \boldsymbol{\tilde{n}}_{\Lambda} \,,
\label{eq:ExactFlowEquationForGammaFinal}
\end{align}
where we have defined the ``single-scale derivative'' $\partial \!\!\!/_{\Lambda}$ in the third line, which acts on the regulator only. This equation is the desired exact flow equation for the effective action of a Fermi-Bose theory in the nonequilibrium Keldysh formalism. Despite its apparent simplicity it is a highly complicated nonlinear functional integro-differential equation, which captures all of the nonperturbative features of the theory.

\subsection{Vertex expansion}
\label{sec:VertexExpansion}
In practice, the exact flow equation~\eqref{eq:ExactFlowEquationForGammaFinal} is too complex to be solved directly. Instead, one has to resort to approximation schemes.

A particularly crude approximation scheme is to neglect the $\Lambda$-dependence of $\boldsymbol{\hat{\Gamma}}_{\Lambda}^{(2)}$ on the right hand side of the flow equation~\eqref{eq:ExactFlowEquationForGammaFinal} and replace it by its initial value at the scale $\Lambda = \Lambda_0$. In this approximation the single-scale derivative $\partial \!\!\!/_{\Lambda}$ turns into an ordinary one and the flow equation can be integrated exactly. For certain cutoff schemes (see Ref.~\onlinecite{BergesTetradisWetterich2002}) this approximation then immediately yields the effective action to one-loop order in perturbation theory
\begin{equation}
\Gamma_{\textrm{1-loop}}[\boldsymbol{\psi}, \boldsymbol{\phi}] = S[\boldsymbol{\psi}, \boldsymbol{\phi}] + \frac{i}{2} \textrm{STr} \, \textrm{ln} \left( \boldsymbol{\hat{S}}^{(2)}[\boldsymbol{\psi}, \boldsymbol{\phi}] \right) \,.
\label{eq:OneLoopEffectiveAction}
\end{equation}
Other approximations, such as the random phase approximation, can be obtained by similar considerations.

In recent years there have been many proposals for systematic approximations of the effective action, which are capable of describing truly nonperturbative phenomena.\cite{BergesTetradisWetterich2002} We here pursue an expansion into powers of fields $\boldsymbol{\Psi}$ and $\boldsymbol{\phi}$, following Refs.~ \onlinecite{MetznerHonercampEtal2012, KopietzBook, SchuetzKopietz2006}. Assuming the effective action to be an analytic functional of the fields, we can perform a formally exact Taylor expansion, known as ``vertex expansion''. It can be employed to replace the single functional integro-differential equation by an equivalent infinite hierarchy of coupled ordinary integro-differential equations for the one-particle irreducible vertex functions. Clearly, to solve the complete hierarchy exactly is still an impossible task. However, a truncation of the infinite hierarchy at a certain finite order is still nonperturbative in essence and does not necessarily rely on the presence of a smallness parameter in the interaction $S_{\textrm{int}}$.

Taking into account that the bosonic field may develop a finite expectation value $\bar{\phi}_c(x)$, \textit{e.g.} due to a finite external scalar potential, we should expand the bosonic field around this macroscopic field, rather than around zero,
\begin{equation}
\phi_c(x) = \bar{\phi}_c(x) + \Delta \phi_c(x) \,, \quad \phi_q(x) = \Delta \phi_q(x)
\label{eq:ExpansionPhi}
\end{equation}
The general vertex expansion in the presence of bosonic field expectation values has been worked out for ``superfields'', a condensed notation collecting fermionic and bosonic degrees of freedom into a single field, in thermal equilibrium by Sch\"utz and Kopietz.\cite{SchuetzKopietz2006, KopietzBook} In our case the vertex expansion reads
\begin{widetext}

\begin{align}
\Gamma_{\Lambda}[\boldsymbol{\psi}, \boldsymbol{\phi}] =& \sum_{m = 0}^{\infty} \sum_{n = 0}^{\infty} \frac{(-1)^m}{(m!)^2} \frac{1}{n!} \int_{x_m, x_m'} \sum_{i_m, i_m'} \sum_{\alpha_m, \alpha_m'} \int_{y_n} \sum_{\beta_n} \nonumber \\
&\times \Gamma_{\Lambda}^{(2m, n)}(x_1 i_1 \alpha_1, \ldots, x_m i_m \alpha_m; x_{1}' i_{1}' \alpha_{1}', \ldots, x_{m}' i_{m}' \alpha_{m}'; y_1 \beta_1, \ldots, y_n \beta_n) \nonumber \\
&\times\psi_{i_1 \alpha_1}^{\dagger}(x_1) \cdots \psi_{i_m \alpha_m}^{\dagger}(x_m) \psi_{i_m' \alpha_m'}(x_m') \cdots \psi_{i_1' \alpha_1'}(x_1') \Delta \phi_{\beta_1}(y_1) \cdots \Delta \phi_{\beta_n}(y_n) \,.
\label{eq:VertexExpansion}
\end{align}
Here, latin indices $i_n$ collectively denote the discrete fermionic degrees of freedom, sublattice, valley and spin, whereas greek indices $\alpha_n, \beta_n$ are reserved for the degrees of freedom in Keldysh space, the classical and quantum components. The coefficient-functions $\Gamma_{\Lambda}^{(2m, n)}$ in this expansion define the one-particle irreducible vertex functions
\begin{align}
&\Gamma_{\Lambda}^{(2m, n)}(x_1 i_1 \alpha_1, \ldots, x_m i_m \alpha_m; x_{1}' i_{1}' \alpha_{1}', \ldots, x_{m}' i_{m}' \alpha_{m}'; y_1 \beta_1, \ldots, y_n \beta_n) = \nonumber \\
&\frac{\delta^{(2m + n)} \Gamma_{\Lambda}}{\delta \psi_{i_1 \alpha_1}^{\dagger}(x_1) \cdots \delta \psi_{i_m \alpha_m}^{\dagger}(x_m) \delta \psi_{i_m' \alpha_m'}(x_m') \cdots \delta \psi_{i_1' \alpha_1'}(x_1') \delta \phi_{\beta_1}(y_1) \cdots \delta \phi_{\beta_n}(y_n)} \Bigg|_{\phi_c = \bar{\phi}_c} \,.
\label{eq:DefinitionVertexFunctions}
\end{align}
In the above definition it is understood that after performing the $(2m + n)$-fold derivative, all fields have to be set to zero except the classical component of the bosonic field, which is set to its possibly nonzero expectation value $\bar{\phi}_c(x)$. This notation has already been employed in Eqs.~\eqref{eq:SelfenergyDefinitionFermions} and~\eqref{eq:SelfenergyDefinitionBoson}. Further, the normalization of the partition function implies that vertex functions which possess classical indices only vanish identically.\cite{ChouSuYu1985}

To obtain the hierarchy of flow equations for the vertex functions, we have to insert the expansion \eqref{eq:VertexExpansion} into the exact flow equation \eqref{eq:ExactFlowEquationForGammaFinal} and compare coefficients. It is important to emphasize that both the vertex functions, as well as the bosonic expectation value are functions of the flowing cutoff $\Lambda$. Thus, we obtain two contributions on the left hand side of \eqref{eq:ExactFlowEquationForGammaFinal}
\begin{align}
\partial_{\Lambda} \Gamma_{\Lambda} =& \sum_{m = 0}^{\infty} \sum_{n = 0}^{\infty} \frac{(-1)^m}{(m!)^2} \frac{1}{n!} \int_{x_m, x_m'} \sum_{i_m, i_m'} \sum_{\alpha_m, \alpha_m'} \int_{y_n} \sum_{\beta_n} \nonumber \\
&\times \left[ \partial_{\Lambda} \Gamma_{\Lambda}^{(2m, n)}(\ldots; y_1 \beta_1, \ldots, y_n \beta_n) - \int_y \Gamma_{\Lambda}^{(2m, n + 1)}(\ldots; y_1 \beta_1, \ldots, y_n \beta_n, y c) \partial_{\Lambda} \bar{\phi}_c(y) \right] \nonumber \\
&\times\psi_{i_1 \alpha_1}^{\dagger}(x_1) \cdots \psi_{i_m \alpha_m}^{\dagger}(x_m) \psi_{i_m' \alpha_m'}(x_m') \cdots \psi_{i_1' \alpha_1'}(x_1') \Delta \phi_{\beta_1}(y_1) \cdots \Delta \phi_{\beta_n}(y_n) \,.
\label{eq:LambdaDerivativeOfVertexExpansion}
\end{align}
%
%
In the second line we suppressed the fermionic arguments of the vertex functions $\Gamma_{\Lambda}^{(2m, n)}$ and $\Gamma_{\Lambda}^{(2m, n + 1)}$ for clarity. For the right hand side it is beneficial to separate the field-independent part from the field-dependent part of $\boldsymbol{\hat{\Gamma}}_{\Lambda}^{(2)}$. Recalling the generalized Dyson equation \eqref{eq:GeneralizedDysonEquation}, we write~\cite{MetznerHonercampEtal2012}
\begin{equation}
\boldsymbol{\hat{\Gamma}}_{\Lambda}^{(2)}[\boldsymbol{\psi}, \boldsymbol{\phi}] = \boldsymbol{\hat{\mathcal{G}}}_{\Lambda}^{-1} - \boldsymbol{\hat{\Sigma}}_{\Lambda}[\boldsymbol{\psi}, \boldsymbol{\phi}] \,,
\label{eq:SecondFunctionalDerivativeMatrixGammaRewritten}
\end{equation}
with
\begin{equation}
\boldsymbol{\hat{\mathcal{G}}}_{\Lambda}^{-1} = \boldsymbol{\hat{\Gamma}}_{\Lambda}^{(2)}|_{\phi_c = \bar{\phi}_c} \,, \quad
\boldsymbol{\hat{\Sigma}}_{\Lambda}[\boldsymbol{\psi}, \boldsymbol{\phi}] = \boldsymbol{\hat{\Gamma}}_{\Lambda}^{(2)}|_{\phi_c = \bar{\phi}_c} - \boldsymbol{\hat{\Gamma}}_{\Lambda}^{(2)} \,.
\label{eq:FullInversePropagatorMatrixFieldDependentSelfEnergyMatrix}
\end{equation}
Here $\boldsymbol{\hat{\mathcal{G}}}_{\Lambda}^{-1}$ is a $3 \times 3$ matrix in field space, which contains the unregularized inverse full propagators, see Eqs.~\eqref{eq:SelfenergyDefinitionFermions} and \eqref{eq:SelfenergyDefinitionBoson}, whereas $\boldsymbol{\hat{\Sigma}}_{\Lambda}[\boldsymbol{\psi}, \boldsymbol{\phi}]$ is the field dependent self-energy, which must not be confused with the (field independent) self-energy in the inverse full propagators. Now we can expand the logarithm on the right hand side of Eq.~\eqref{eq:ExactFlowEquationForGammaFinal} in terms of (regularized) full propagators $\left( \boldsymbol{\hat{\mathcal{G}}}_{\Lambda}^{-1} + \boldsymbol{\hat{\mathcal{R}}}_{\Lambda} \right)^{-1}$ as follows
\begin{align}
\textrm{ln} \left( \boldsymbol{\hat{\Gamma}}_{\Lambda}^{(2)} + \boldsymbol{\hat{\mathcal{R}}}_{\Lambda} \right) &= \textrm{ln} \left( 1 - \left( \boldsymbol{\hat{\mathcal{G}}}_{\Lambda}^{-1} + \boldsymbol{\hat{\mathcal{R}}}_{\Lambda} \right)^{-1} \boldsymbol{\hat{\Sigma}}_{\Lambda}[\boldsymbol{\psi}, \boldsymbol{\phi}] \right) \nonumber \\
&= - \sum_{n = 1}^{\infty} \frac{1}{n} \left( \left( \boldsymbol{\hat{\mathcal{G}}}_{\Lambda}^{-1} + \boldsymbol{\hat{\mathcal{R}}}_{\Lambda} \right)^{-1} \boldsymbol{\hat{\Sigma}}_{\Lambda}[\boldsymbol{\psi}, \boldsymbol{\phi}] \right)^n \,.
\label{eq:ExpansionRHSOfExactFlowEquation}
\end{align}
The desired hierarchy of flow equations is given by comparing coefficients in the expansions of \eqref{eq:LambdaDerivativeOfVertexExpansion} and \eqref{eq:ExpansionRHSOfExactFlowEquation}. This can be done in a systematic way, because, by construction, the field dependent self-energy $\boldsymbol{\hat{\Sigma}}_{\Lambda}[\boldsymbol{\psi}, \boldsymbol{\phi}]$ only contains terms which are at least linear in one field variable.

In the following we present a truncated set of equations, with the further approximation that only vertices with one bosonic and two fermionic legs have been kept. The motivation for this approximation is, that these structures are already present in the bare action. Accounting for the purely bosonic three-vertex and other higher order vertices, which are inevitably generated by the flow, is possible by using the strategy explained above. The equation for the bosonic field expectation value reads
\begin{align}
\int\nolimits_{x_1'} \Big( (V^{-1} + \Pi_{\Lambda}^R + R_{\textrm{b}, \Lambda}^R)&(x_1, x_1') \partial_{\Lambda} \bar{\phi}_c(x_1') + \partial_{\Lambda} R_{\textrm{b}, \Lambda}^R(x_1, x_1') \bar{\phi}_c(x_1') \Big) \nonumber \\
&= - \frac{i}{2} \partial \!\!\!/_{\Lambda} \sum_{\alpha, \beta} \sum_{k, l} \int\nolimits_{x_1', x_2'} G_{\Lambda, kl}^{\alpha \beta}(x_1', x_2') \Gamma_{\Lambda, lk}^{(2, 1)}(x_2' \beta, x_1' \alpha; x_1 q) - \partial_{\Lambda} \tilde{n}_{\Lambda}(x_1) \,,
\label{eq:FlowEqFieldExpectationValue}
\end{align}
\end{widetext}
The derivation of this equation makes use of the equation of motion \eqref{eq:EquationsOfMotionGamma3} at its extremal value $\boldsymbol{\Psi} = 0, \boldsymbol{\Delta \phi} = 0$, replacing the flow of the one-point function.

In the presence of a bosonic regulator a graphical representation of the above equation is rather exceptional and not very helpful. However, for purely fermionic cutoff schemes, we recognize the typical tadpole structure, also known as Hartree diagrams, by applying $(V^{-1} + \Pi_{\Lambda}^R)^{-1}$ on both sides of the equation. Further note the counterterm flow on the right hand side. It is the only location where the background charge density $\tilde{n}(x)$ enters the flow equations explicitly. We can understand its presence here by considering exemplarily the space-time translation invariant system at finite density. In that case the first term on the right hand side is finite and closely related to the charge carrier density. (In fact, in the simple truncation scheme where the three-vertex flow is neglected it is identical to the charge carrier density.) In turn, this would imply that the expectation value $\bar{\phi}_c$ has to be finite. The counterterm, however, cancels the finite contribution on the right hand side at any scale, such that the expectation value is consistently removed from the theory and all tadpole diagrams with it. In other physical situations, depending on the experimental setup, the counterterm flow has to be constructed by further physical considerations.

We here show the flow equations for the fermionic self-energy, bosonic polarization and the three-vertex in their graphical form only. Their explicit analytical form is given in App.~\ref{sec:AnalyticalFormOfTheVertexFlowEquations},
\begin{align}
\partial_{\Lambda} \boldsymbol{\hat{\Sigma}}_{\Lambda} &= i \partial \!\!\!/_{\Lambda}
\raisebox{-1.3cm}{
\scalebox{.8}{
	\begin{fmfgraph}(40,25)
		\fmfleft{i}
		\fmfright{o}
		\fmf{fermion,tension=3}{G1,i}
		\fmfrpolyn{filled=15}{G}{3}
		\fmfpolyn{filled=15}{K}{3}
		\fmf{fermion,tension=1.5}{K3,G3}
		\fmf{wiggly,left=.5,tension=0}{G2,K2}
		\fmf{fermion,tension=3}{o,K1}
	\end{fmfgraph}
}} +
\raisebox{-1.3cm}{
\scalebox{.8}{
	\begin{fmfgraph}(20,25)
	\fmfcmd{
    path quadrant, q[], otimes;
    quadrant = (0, 0) -- (0.5, 0) & quartercircle & (0, 0.5) -- (0, 0);
    for i=1 upto 4: q[i] = quadrant rotated (45 + 90*i); endfor
    otimes = q[1] & q[2] & q[3] & q[4] -- cycle;
		}
	\fmfwizard
		\fmfleft{i}
		\fmfright{o}
		\fmftop{t}
		\fmf{fermion,tension=3}{G1,i}
		\fmfrpolyn{filled=15}{G}{3}
		\fmf{fermion,tension=3}{o,G3}
		\fmffreeze
		\fmf{plain,tension=1}{G2,v1}
		\fmfv{d.sh=otimes,d.f=empty}{v1}	
		\fmf{phantom,tension=.5}{v1,t}
		\fmfdot{t}
	\end{fmfgraph}
}} \,,
\label{eq:DiagrammaticSelfenergyFlowCompact} \\
\partial_{\Lambda} \boldsymbol{\Pi}_{\Lambda} &= \frac{i}{2} \partial \!\!\!/_{\Lambda}
\scalebox{.8}{
	\begin{fmfgraph}(45,2.5)
		\fmfleft{i}
		\fmfright{o}
		\fmf{wiggly,tension=3}{i,G1}
		\fmfrpolyn{filled=15}{G}{3}
		\fmfpolyn{filled=15}{K}{3}
		\fmf{fermion,left=.3,tension=.5}{G2,K2}
		\fmf{fermion,left=.3,tension=.5}{K3,G3}
		\fmf{wiggly,tension=3}{K1,o}
	\end{fmfgraph}
} \,,
\label{eq:DiagrammaticPolarizationFlowCompact}
\end{align}

\begin{align}
\partial_{\Lambda} \boldsymbol{\Gamma}_{\Lambda}^{(2,1)} = i \partial \!\!\!/_{\Lambda}
\raisebox{-.9cm}{
\scalebox{.8}{
	\begin{fmfgraph}(25,25)
		\fmfleft{i1}
		\fmfright{o2,o3}
		\fmfpolyn{filled=15}{G}{3}
		\fmfpolyn{filled=15}{K}{3}
		\fmfpolyn{filled=15}{J}{3}
		\fmf{wiggly,tension=4}{G1,i1}
		\fmf{fermion,tension=4}{o2,K2}
		\fmf{fermion,tension=4}{J3,o3}
		\fmf{wiggly,tension=1}{K3,J2}
		\fmf{fermion,tension=1}{G3,J1}
		\fmf{fermion,tension=1}{K1,G2}
	\end{fmfgraph}
}} \,.
\label{eq:DiagrammaticVertexFlowCompact}
\end{align}
In these diagrams, the straight line corresponds to a fermionic full propagator, the wiggly line to a bosonic full propagator, the triangle to a vertex and the crossed circle to the bosonic field expectation value. The dot above the crossed circle denotes the scale derivative acting on the expectation value. Summation over discrete degrees of freedom (including Keldysh space) and integration over continuous ones is implied. The above flow equations closely resemble one-loop perturbation theory, a fact which is not surprising, since the exact flow equation~\eqref{eq:ExactFlowEquationForGammaFinal} has a one-loop structure.

By construction, the single-scale derivative $\partial \!\!\!/_{\Lambda}$ appearing in the above expressions does not act on the vertex functions, but only on the regulator occuring in the expressions for the internal full propagators (such as the factor $G_{\Lambda, kl}^{\alpha \beta}$ in Eq.~\eqref{eq:FlowEqFieldExpectationValue}). In other words, $\partial \!\!\!/_{\Lambda}$ is a scale-derivative at constant self-energy, which yield what is known in the literature as single-scale propagators~\cite{MetznerHonercampEtal2012, KopietzBook}
\begin{align}
\partial \!\!\!/_{\Lambda} \boldsymbol{\hat{G}}_{\Lambda} &= - \boldsymbol{\hat{G}}_{\Lambda} \partial_{\Lambda} \boldsymbol{\hat{R}}_{\textrm{f}, \Lambda} \boldsymbol{\hat{G}}_{\Lambda} \,\,\,\; \equiv \boldsymbol{\hat{S}}_{\textrm{f}, \Lambda} \,, \\
\partial \!\!\!/_{\Lambda} \boldsymbol{D}_{\Lambda} &= - \boldsymbol{D}_{\Lambda} \partial_{\Lambda} 2 \boldsymbol{R}_{\textrm{b}, \Lambda} \boldsymbol{D}_{\Lambda} \equiv \boldsymbol{S}_{\textrm{b}, \Lambda} \,.
\label{eq:SingleScalePropagators}
\end{align}
Graphically, the single-scale propagators are often depicted as a (straight or wiggly) line with a slash. They have the same trigonal structure as the flowing propagators
\begin{equation}
\boldsymbol{\hat{S}}_{\textrm{f}, \Lambda} = \begin{pmatrix} \hat{S}_{\textrm{f}, \Lambda}^K & \hat{S}_{\textrm{f}, \Lambda}^R \\ \hat{S}_{\textrm{f}, \Lambda}^A & 0 \end{pmatrix} \,, \quad \boldsymbol{S}_{\textrm{b}, \Lambda} = \begin{pmatrix} S_{\textrm{b}, \Lambda}^K & S_{\textrm{b}, \Lambda}^R \\ S_{\textrm{b}, \Lambda}^A & 0 \end{pmatrix} \,.
\label{eq:KeldyshStructureSingleScalePropagators}
\end{equation}
The advantage of using the single-scale derivative is, that the computational effort to arrive at the vertex flow equations as well as their analysis is greatly reduced. The reason being, in particular, that $\partial \!\!\!/_{\Lambda}$ obeys the product rule for differentiation, according to which, at the graphical level, for each internal line on the right hand side of Eqs.~\eqref{eq:DiagrammaticSelfenergyFlowCompact}--\eqref{eq:DiagrammaticVertexFlowCompact} the single-scale derivative produces an additional equivalent term, where the corresponding line has been substituted by a single-scale propagator. Therefore, one may perform all analytical manipulations within the integrals first and apply the scale derivative afterwards.

We close this section by discussing the role of correlated initial states in the above set of exact flow equations. Recall from section \ref{sec:ContourTimeGeneratingFunctional} that correlated initial states manifest themselves as higher order terms in the expansion of the correlation functional $K_{\rho}[\psi]$, see Eq.~\eqref{eq:InitialCorrelationFunctional}. The kernels of this expansion would appear within the effective action $\Gamma_{\Lambda}$ as a contribution to the respective higher order vertex function in the expansion~\eqref{eq:VertexExpansion} already at the initial scale $\Lambda_0$.\cite{BergesCox2000} Since a common truncation strategy of the infinite hierarchy of flow equations is to keep only those vertices which are already present in the bare action, the number of flow equations, which should be considered for correlated initial states, grows rapidly. Even for the simplest possible nongaussian extension, which is a quartic term in the fermionic correlation functional, the analysis is considerably impeded. First, one would have to keep the four-vertex contribution to the fermionic self-energy flow, and second it should be revised, if it is justifiable to neglect the four-vertex flow entirely or if at least the flow of some dominant interaction channel has to be taken into account. Owing to the complicated structure of the flow equations, it becomes clear that the study of nongaussian initial correlations is practically limited to a low order.\cite{Note5}
On the other hand the field is vastly unexplored and may lead to interesting new physical effects. In any case the nonequilibrium functional renormalization group as we presented above is an excellent framework for such an undertaking.

\section{Thermal Equilibrium}
\label{sec:ThermalEquilibrium}
As a first application and a test of the methods developed in the previous section, we now apply the general nonequilibrium formalism to the equilibrium case and show how the results of the Matsubara imaginary-time formalism are recovered. In thermal equilibrium physical observables do not depend on time. In particular, the reference time $t_0$ drops out in any calculation, so that the limit $t_0 \rightarrow -\infty$ may be taken at the beginning of the calculation and a Fourier transform to frequency space can be performed. In contrast to the Matsubara formalism, the frequencies in the Keldysh formulation are real and continuous, which removes the need for an analytical continuation at the end of a calculation. The temperature dependence enters through the solution of the kinetic equations and the fluctuation-dissipation theorem, which will be discussed below.

In the following we further restrict ourselves to spatially translation invariant systems, setting the external electromagnetic potentials to zero. Since the propagators and each vertex now conserve energy and momentum, the flow equations simplify considerably. We also limit ourselves to intrinsic, freestanding graphene, setting the chemical potential $\mu$ and the background charge density $\tilde{n}$ to zero, and the dielectric constant of the medium $\epsilon_0$ to unity. As a consequence the bosonic field expectation value and the counterterm vanish. After discussing some general aspects, we present a simple truncation scheme for the flow equations, and solve the resulting system of equations numerically for finite temperatures.

\subsection{Fluctuation-dissipation theorem and cutoff schemes}
\label{sec:FluctuationDissipationTheoremAndCutoffSchemes}
The equilibrium state is uniquely specified by the Boltzmann statistical operator $\hat{\rho} = \textrm{exp}(- \beta \hat{H})$. This particular density matrix leads to a periodicity of the fermionic and bosonic field operators along the imaginary time axis, which can be expressed by the KMS boundary conditions.\cite{NegeleOrlandBook} Eventually, these boundary conditions manifest themselves as constraining relations between the various $n$-point correlation functions, which is known as the fluctuation-dissipation theorem. Demanding its validity at any scale greatly reduces the numerical effort, since the flow equations themselves have to preserve these constraints. Thus, the number of independent flow equations is diminished. We here concentrate on the fluctuation-dissipation relation for the connected two-point correlators and self-energies. We refer to Refs.~\onlinecite{CarringtonETAL2007} and \onlinecite{SiebererETAL2015} for further reading.

In the Keldysh formalism the fluctuation-dissipation theorem can be very elegantly formulated. The necessary condition for thermal equilibrium is the vanishing of the kinetic term in the quantum kinetic equations \eqref{eq:QuantumKineticEquationFermions} and~\eqref{eq:QuantumKineticEquationBoson}. Assuming that the hermitian matrix $\hat{\mathcal{F}}_{\Lambda}(\vec{k}, \varepsilon)$ is proportional to the unit matrix we thus have
\begin{align}
\hat{\Sigma}_{\Lambda}^K(\vec{k}, \varepsilon) &= \mathcal{F}_{\Lambda}(\vec{k}, \varepsilon) \left( \hat{\Sigma}_{\Lambda}^R(\vec{k}, \varepsilon) - \hat{\Sigma}_{\Lambda}^A(\vec{k}, \varepsilon) \right) \,,
\label{eq:FluctuationDissipationTheoremSelfEnergy} \\
\Pi_{\Lambda}^K(\vec{q}, \omega) &= \mathcal{B}_{\Lambda}(\vec{q}, \omega) \left( \Pi_{\Lambda}^R(\vec{q}, \omega) - \Pi_{\Lambda}^A(\vec{q}, \omega) \right) \,.
\label{eq:FluctuationDissipationTheoremPolarization}
\end{align}
The fluctuation-dissipation theorem states that the distribution functions $\mathcal{F}_{\Lambda}$ and $\mathcal{B}_{\Lambda}$ take the simple, scale independent form
\begin{align}
\mathcal{F}_{\Lambda}(\vec{k}, \varepsilon) &= \textrm{tanh}\left( \frac{\varepsilon}{2T} \right) \,, \\
\mathcal{B}_{\Lambda}(\vec{q}, \omega) &= \textrm{coth}\left( \frac{\omega}{2T} \right) \,.
\label{eq:EquilibriumDistributionFunctions}
\end{align}
Since the equilibrium solution is unique, their independence of the scale $\Lambda$ is crucial. Using the above solution, we can immediately write down the corresponding Keldysh propagators
\begin{align}
\hat{G}_{\Lambda}^K(\vec{k}, \varepsilon) &= \textrm{tanh}\left( \frac{\varepsilon}{2T} \right) \left( \hat{G}_{\Lambda}^R(\vec{k}, \varepsilon) - \hat{G}_{\Lambda}^A(\vec{k}, \varepsilon) \right) \,,
\label{eq:EquilibriumKeldyshPropagatorFermions} \\
D_{\Lambda}^K(\vec{q}, \omega) &= \textrm{coth}\left( \frac{\omega}{2T} \right) \Big( D_{\Lambda}^R(\vec{q}, \omega) - D_{\Lambda}^A(\vec{q}, \omega) \Big) \,.
\label{eq:EquilibriumKeldyshPropagatorBoson}
\end{align}

Whereas the fluctuation-dissipation theorem is generally valid in thermal equilibrium for the physical limit $\Lambda \to 0$, its validity at all scales $\Lambda$ is not automatic. Requiring Eqs.~\eqref{eq:EquilibriumDistributionFunctions} for arbitrary cutoff $\Lambda$ puts strong constraints on the choice of the infrared regulators. As discussed in the previous section, these constraints have to be implemented together with the restrictions that ensure that the cutoff scheme preserves causality and respects all the symmetries of the model. 

Following Ref.~\onlinecite{BauerKopietz2015}, we now describe a regularization scheme that meets these conditions. We have adopted this regularization scheme for our numerical calculations, in order to facilitate the comparison of our results and those of Ref.\ \onlinecite{BauerKopietz2015}. In this scheme, regularization is applied in the fermionic sector only, 
\begin{equation}
  R_{\textrm{b}, \Lambda}^{R/A/K} = 0.
\end{equation}
As a consequence, the bosonic single-scale propagators vanish identically. For the fermionic degrees of freedom, we consider a regulator with momentum dependence only, 
\begin{align}
&\hat{R}_{\textrm{f}, \Lambda}^R(\vec{k}, \varepsilon) = \hat{R}_{\textrm{f}, \Lambda}^A(\vec{k}, \varepsilon) = \hat{G}_{0, \Lambda}^{-1}(\vec{k}, \varepsilon) - \hat{G}_0^{-1}(\vec{k}, \varepsilon) \,, \nonumber \\
&\hat{R}_{\textrm{b}, \Lambda}^K(\vec{k}, \varepsilon) = 0
\label{eq:ChoiceOfCutoffScheme}
\end{align}
with
\begin{equation}
\hat{G}_{0, \Lambda}^{-1}(\vec{k}, \varepsilon) = \hat{G}_0^{-1}(\vec{k}, \varepsilon) (\Theta(k - \Lambda))^{-1} \,.
\label{eq:SharpMomentumCutoff}
\end{equation}
The absence of a frequency dependence of the regulator function implies that the frequency structure of the propagators is untouched by the regularization procedure and causality is manifestly preserved. The sharp $\Theta$-function cutoff in momentum space simplifies the flow equations even further by eliminating one of the integrations involved on their right hand side. 

Several aspects of this choice of the regulator function are worthwhile discussing. The first issue is the role of the Fermi surface. At charge neutrality the Fermi surface consists of the points located at the $K_+$ and $K_-$ points, a fact that is not altered by the interaction. This is a major simplification, because there is no need to adapt the regulators to a continuously changing Fermi surface. Since this simplification is special to the charge neutrality point, other regularization schemes may be preferable away from it, see our discussion below.

Second, the above choice of regularization function transforms the additive regularization into a multiplicative one. Such multiplicative regularizations are also common in the literature, see, {\em e.g.}, Refs.~\onlinecite{MetznerHonercampEtal2012, KopietzBook}. The Keldysh regulator has been set to zero in order to guarantee the trivial initial conditions $\Gamma_{\Lambda_0} = S$. Although now the kinetic terms in the kinetic equations contain explicitly the regulators, it is still possible to obtain the scale independent equilibrium solutions of the previous section. This fact is a simple consequence of the scalar multiplicative cutoff. 

At the end of section~\ref{sec:DysonAndQuantumKineticEquationsInTheFunctionalRenormalizationGroup} we discussed that a parametrization of the Keldysh regulators $\hat{R}_{\textrm{f}, \Lambda}^K$ and $R_{\textrm{b}, \Lambda}^K$ in terms of the distribution functions $\hat{\mathcal{F}}_{\Lambda}$ and $\mathcal{B}_{\Lambda}$, respectively, in principle leads to a simplification of the kinetic terms in the kinetic equations, see Eqs.~\eqref{eq:QuantumKineticEquationFermions}--\eqref{eq:ParametrizationKeldyshRegulatorsBoson}. In this parametrization the kinetic terms no longer explicitly contain the regularization functions. As a result the kinetic equations can be solved immediately by the above scale independent distribution functions. This fact applies to regulator functions that act in the momentum and/or frequency domain. The possibility to use cutoffs in the frequency domain that manifestly preserve causality is a major technical advantage of the Keldysh formulation and does not exist for frequency cutoffs in the imaginary-time formulation, where the causality structure is usually destroyed. Of course, in frequency-independent regularization schemes, such as the one of Eqs.~\eqref{eq:ChoiceOfCutoffScheme}, causality issues are avoided for both approaches. An additional advantage of a frequency cutoff in the fermionic sector is that no explicit reference to a Fermi surface needs to be made. 

An example for a cutoff scheme, which incorporates all of the above mentioned properties, is the ``hybridization cutoff'' of Jakobs \textit{et~al.}\cite{JakobsPletyukovSchoeller2010, JakobsMedenSchoeller2007} In this scheme the infinitesimal regulators $\pm i0$ in the inverse bare propagators and the Keldysh blocks are elevated to cutoff dependent quantities $\pm i \Lambda$. Being essentially a frequency cutoff, the hybridization scheme is particularly useful in those cases where a momentum cutoff is not appropriate, such as graphene away from the charge neutrality point or the presence of a finite magnetic field. In both cases, the Fermi surface (if it can be defined at all) will be subject to change during the renormalization group flow, requiring a continuous adjustment of the momentum cutoff. The frequency cutoff of Refs.~\onlinecite{JakobsPletyukovSchoeller2010, JakobsMedenSchoeller2007}, on the other hand, is insensitive to a changing Fermi surface and compatible with spatially varying external fields. Another example of a frequency cutoff is the ``outscattering rate cutoff'' employed by Kloss and Kopietz.\cite{KlossKopietz2011} It is similar to the hybridization cutoff, but has the important difference that the Keldysh blocks of the inverse free propagators are not regularized. In this case the distribution functions become explicitly scale dependent and the fluctuation dissipation theorem is manifestly violated, making the outscattering rate cutoff not suitable for an equilibrium setting.

\subsection{Dressed flowing propagators}
\label{sec:FlowingPropagators}
After having discussed the regularization scheme, we can now give explicit expressions for the dressed flowing propagators, which are central to the flow equations of the functional renormalization group. The temperature arguments of the fermionic and bosonic self-energies are suppressed in the following.

As discussed in Sec.~\ref{sec:GreenFunctions}, the expressions for the fermionic propagators take their simplest form in the chiral basis. Since by construction of the regulators the exact flow equation preserves chirality at all scales, the same holds true for the fermionic self-energy and the flowing propagators
\begin{align}
\hat{\Sigma}_{\Lambda}^{R/A}(\vec{k}, \varepsilon) &= \sum_{\pm} \hat{\mathcal{P}}_{\pm}(\hat{k}) \Sigma_{\pm, \Lambda}^{R/A}(k, \varepsilon) \,,
\label{eq:SelfenergyChiralBasis} \\
\hat{G}_{\Lambda}^{R/A}(\vec{k}, \varepsilon) &= \sum_{\pm} \hat{\mathcal{P}}_{\pm}(\hat{k}) G_{\pm, \Lambda}^{R/A}(k, \varepsilon)\,,
\label{eq:DressedFlowingFermionRAPropagatorsChiralDecomposition}
\end{align}
%
%
where the $\hat{\mathcal{P}}_{\pm}(\hat{k})$ are the chiral projection operators, see Eq.~\eqref{eq:ChiralProjectors}. Thus, the retarded and advanced chiral flowing propagators can be written in the compact form
\begin{align}
G_{\pm, \Lambda}^{R/A}(k, \varepsilon) &= \frac{\Theta(k - \Lambda)}{\varepsilon \mp v_F k - \Sigma_{\pm, \Lambda}^{R/A}(k, \varepsilon)} \,, 
\label{eq:DressedFlowingChiralRAPropagators} \\
&= \frac{\Theta(k - \Lambda)}{ \left( \varepsilon - \Sigma_{\varepsilon, \Lambda}^{R/A}(k, \varepsilon) \right) \mp \left( v_F + \Sigma_{v, \Lambda}^{R/A}(k, \varepsilon) \right) k} \,, \nonumber
\end{align}
where we have defined
\begin{subequations}
\begin{align}
\Sigma_{\varepsilon, \Lambda}^{R/A}(k, \varepsilon) &= \frac{1}{2} \left( \Sigma_{+, \Lambda}^{R/A} + \Sigma_{-, \Lambda}^{R/A} \right)(k, \varepsilon) \,,
\label{eq:ScalarSelfenergiesA} \\
\Sigma_{v, \Lambda}^{R/A}(k, \varepsilon) &= \frac{1}{2 k} \left( \Sigma_{+, \Lambda}^{R/A} - \Sigma_{-, \Lambda}^{R/A} \right)(k, \varepsilon) \,.
\label{eq:ScalarSelfenergiesB}
\end{align}
\end{subequations}
Recall that the single-scale derivative only acts on the $\Theta$-function, such that the sharp momentum cutoff yields a particularly simple single-scale propagator.

The retarded and advanced propagators in the bosonic sector are given by
\begin{equation}
D_{\Lambda}^{R/A}(q, \omega) = \frac{1}{2} \frac{1}{V^{-1}(q) + \Pi_{\Lambda}^{R/A}(q, \omega)} \,,\vspace{1mm}
\label{eq:DressedFlowingBosonRAPropagators}
\end{equation}
where $V(q)$ is the two-dimensional Fourier transform of the Coulomb interaction,
\begin{align}
V(q) = \frac{2 \pi e^2}{q} \,.
\label{eq:CoulombInteractionFourier}
\end{align}
The $\Lambda$ dependence of the bosonic propagators is entirely determined by the flowing polarization function. By introducing the dielectric function
\begin{equation}
\epsilon_{\Lambda}^{R/A}(q, \omega) \equiv 1 + V(q) \Pi_{\Lambda}^{R/A}(q, \omega) \,,
\label{eq:DefinitionDielectricFunction}
\end{equation}
the propagators can be written in the convenient form
\begin{equation}
D_{\Lambda}^{R/A}(q, \omega) = \frac{1}{2} \frac{V(q)}{\epsilon_{\Lambda}^{R/A}(q, \omega)} \,.
\label{eq:DressedFlowingBosonRAPropagatorsDielectricFunction}
\end{equation}

\subsection{Fermi velocity and static dielectric function at finite temperature}
\label{sec:FermiVelocityAndStaticDielectricFunctionAtFiniteTemperature}
We now proceed to solve the truncated flow equations using a finite-temperature real-time analogue of the truncation scheme employed by Bauer \textit{et al.}\cite{BauerKopietz2015} We consider intrinsic graphene, so that the bosonic field expectation value $\phi_c$ and the counterterm are absent. The system of equations~\eqref{eq:DiagrammaticSelfenergyFlowCompact}--\eqref{eq:DiagrammaticVertexFlowCompact}, see also App.~\ref{sec:AnalyticalFormOfTheVertexFlowEquations}, is further simplified by neglecting the flow of the three-vertex functions entirely, keeping these at their initial values at $\Lambda = \Lambda_0$. We also neglect the $\Lambda$ dependence of the scalar self energy $\Sigma_{\varepsilon, \Lambda}^{R/A}$ of Eq.~\eqref{eq:ScalarSelfenergiesA}, as well as the frequency dependence of the scalar self energy $\Sigma_{v, \Lambda}^{R/A}$ of Eq.~\eqref{eq:ScalarSelfenergiesB}. These approximations lead to well-defined $\Lambda$-dependent poles of the single-particle propagators~\eqref{eq:DressedFlowingChiralRAPropagators} at
\begin{equation}
\xi_{\Lambda}(k) = v_{\Lambda}(k) k \,,
\end{equation}
where the renormalized Fermi velocity is given by 
\begin{equation}
v_{\Lambda}(k) = v_F + \Sigma_{v, \Lambda}(k) \,.
\end{equation}
Finally, we neglect the frequency dependence of the dielectric function $\epsilon_{\Lambda}^{R/A}(q, \omega) = \epsilon_{\Lambda}(q)$. As a consequence the bosonic Keldysh propagator remains identically zero during the flow.

The complete truncation scheme can be conveniently expressed if we parameterize the effective action as
\begin{widetext}

\begin{align}
\Gamma_{\Lambda}[\boldsymbol{\Psi}, \boldsymbol{\phi}] &= \int_{\vec{k}, \varepsilon} \boldsymbol{\Psi^{\dagger}}(\vec{k}, \varepsilon) \sigma_0^s \otimes \begin{pmatrix} 0 & \Sigma_0 (\varepsilon - i0) + v_{\Lambda}(k) \vec{\Sigma} \cdot \vec{k} \\ \Sigma_0 (\varepsilon + i0) + v_{\Lambda}(k) \vec{\Sigma} \cdot \vec{k} & 2i0 \, \textrm{tanh} \left( \frac{\varepsilon}{2T} \right) \Sigma_0 \end{pmatrix} \boldsymbol{\Psi}(\vec{k}, \varepsilon) \nonumber \\[1mm]
&+ \int_{\vec{q}, \omega} \boldsymbol{\phi}^{\intercal}(- \vec{q}, - \omega) \begin{pmatrix} 0 & \big( V(q)/\epsilon_{\Lambda}(q) \big)^{-1} \\ \big( V(q)/\epsilon_{\Lambda}(q) \big)^{-1} & 0 \end{pmatrix} \boldsymbol{\phi}(\vec{q}, \omega) \nonumber \\[1mm]
&- \int_{\vec{k},\varepsilon, \vec{q}, \omega} \boldsymbol{\Psi^{\dagger}}(\vec{k} + \vec{q}, \varepsilon + \omega) \begin{pmatrix} \phi_q(\vec{q}, \omega) & \phi_c(\vec{q}, \omega) \\ \phi_c(\vec{q}, \omega) & \phi_q(\vec{q}, \omega) \end{pmatrix} \boldsymbol{\Psi}(\vec{k}, \varepsilon) \,.
\label{eq:TruncatedEffectiveAction}
\end{align}
%
%
We note that if one wishes to go beyond the static approximation of Eq.~\eqref{eq:TruncatedEffectiveAction}, and include the dynamical effects of plasmons and quasiparticle wavefunction renormalization, one should not neglect the three-vertex flow entirely. A naive extension, where only the renormalization of $\Sigma_{\varepsilon, \Lambda}^{R/A}$ and the frequency dependences of $\Sigma_{v, \Lambda}^{R/A}$ and $\epsilon_{\Lambda}^{R/A}$ are taken into account, is not sufficient. As Bauer \textit{et~al.} have shown,\cite{BauerKopietz2015} one should at least include the marginal part of the three-vertex in the analysis. In that case the vertex flow reduces to a differential form of a Ward identity, leading to a partial cancellation of fermionic self-energy- and vertex-corrections. Neglecting the vertex flow would violate the Ward identity and lead to an inconsistency in the flow of the quasiparticle wavefunction renormalization.

%

The sequence of approximations described above results in two coupled flow equations, one for the Fermi velocity $v_{\Lambda}(k)$ and one for the static dielectric function $\epsilon_{\Lambda}(q)$. The approximations are self consistent in the sense that neither a quasipaticle wavefunction renormalization nor a frequency dependence of the dielectric function are generated during the flow.
Within the truncation of the effective action given above, we obtain the flow equation for the Fermi velocity
\begin{equation}
\Lambda \partial_{\Lambda} v_{\Lambda}(k) = - \frac{e^2}{2 \pi} \frac{\Lambda}{k} \int_0^{\pi} d\varphi \, \textrm{tanh} \left( \frac{\xi_{\Lambda}(\Lambda)}{2 T} \right) \frac{\textrm{cos} \varphi}{\sqrt{1 + \left( \tfrac{k}{\Lambda} \right)^2 - 2 \tfrac{k}{\Lambda} \textrm{cos} \varphi}} \frac{1}{\epsilon_{\Lambda} \left(\Lambda \sqrt{1 + \left( \tfrac{k}{\Lambda} \right)^2 - 2 \tfrac{k}{\Lambda} \textrm{cos} \varphi} \right)} \,,
\label{eq:FlowEqFermiVelocityFiniteT}
\end{equation}
whereas the flow equation for the static dielectric function takes the form
\begin{align}
\Lambda \partial_{\Lambda} \epsilon_{\Lambda}(q) = - \frac{2 e^2}{\pi} q &\int_0^{\pi/2} d\varphi \, \Theta \left(\textrm{cos} \varphi + \frac{2 \Lambda}{q} - 1 \right) \frac{1}{\sqrt{ \left( 1 + \frac{q}{2 \Lambda} \textrm{cos} \varphi \right)^2 - \left( \frac{q}{2 \Lambda} \right)^2}}
\label{eq:FlowEqDielectricFunctionFiniteT} \\
&\times \Bigg[ \left( \textrm{tanh} \left( \frac{\xi_{\Lambda}(\Lambda)}{2 T} \right) + \textrm{tanh} \left( \frac{\xi_{\Lambda}(\Lambda + q \textrm{cos} \varphi)}{2 T} \right) \right) \frac{\textrm{sin}^2 \varphi}{\xi_{\Lambda}(\Lambda) + \xi_{\Lambda}(\Lambda + q \textrm{cos} \varphi)} \nonumber \\
&+ \;\;\, \left( \textrm{tanh} \left( \frac{\xi_{\Lambda}(\Lambda)}{2 T} \right) - \textrm{tanh} \left( \frac{\xi_{\Lambda}(\Lambda + q \textrm{cos} \varphi)}{2 T} \right) \right) \frac{(2 \Lambda/q + \textrm{cos} \varphi)^2 - 1}{\xi_{\Lambda}(\Lambda) - \xi_{\Lambda}(\Lambda + q \textrm{cos} \varphi)} \Bigg] \,. \nonumber
\end{align}

\end{widetext}
The derivation of Eq.~\eqref{eq:FlowEqDielectricFunctionFiniteT} requires the use of elliptic coordinates. At the initial scale $\Lambda = \Lambda_0$ the fermionic and bosonic self-energies vanish, which translates to the initial conditions $v_{\Lambda_0}(k) = v_F, \epsilon_{\Lambda_0}(k) = 1$. In the limit $T \rightarrow 0$ our equations reduce to the expressions given in Ref.~\onlinecite{BauerKopietz2015}. The temperature dependence enters the Fermi velocity flow equation only as a simple factor in the integrand, due to the absence of plasmonic effects. The temperature dependence of the dielectric function flow equation, on the other hand, is more complicated. The two contributions in the second and third line of Eq.~\eqref{eq:FlowEqDielectricFunctionFiniteT} can be traced back to inter- and intra-band transitions, repectively. At $T = 0$ the valence band is fully occupied, while the conduction band is empty. Thus, the fermionic phase space for intra-band transitions is Pauli blocked and only inter-band transitions contribute to the polarization function. A finite temperature, however, lifts this Pauli blockade by opening the intra-band phase space for momenta of the order $T$, leading to the additional term in the third line.

The above equations have been solved numerically for different temperatures with the dimensionless coupling constant $\alpha = e^2/v_F = 2.2$. Specifically, they have been rewritten as pure Volterra integral equations of the second kind by integration over the scale variable $\Lambda$, see Ref.~\onlinecite{WazwazBook}, and using the initial conditions. We discretized the parameter spaces by nonuniform, adaptive grids, which were interpolated linearly when intermediate values were required. The case $k = 0$ could not be included in the grids due to divergent terms. Therefore, we built the grids down to $k/\Lambda_0 = 10^{-5}$ and extrapolated for lower momenta if necessary. The coupled system of integral equations has been solved iteratively, starting from the initial values $v_{\Lambda}(k) = v_{\rm F}$ and $\epsilon_{\Lambda}(q) = 1$ for the zero-temperature calculation and continuing the iteration until a self-consistent solution was obtained. During the iterative procedure the grids were occasionally refined according to a gradient criterion. For finite temperature we used previously computed and converged results at a nearby temperature as an initial value in order to minimize the computation time. 

\begin{figure}
	\centering
		\includegraphics[width=0.45\textwidth]{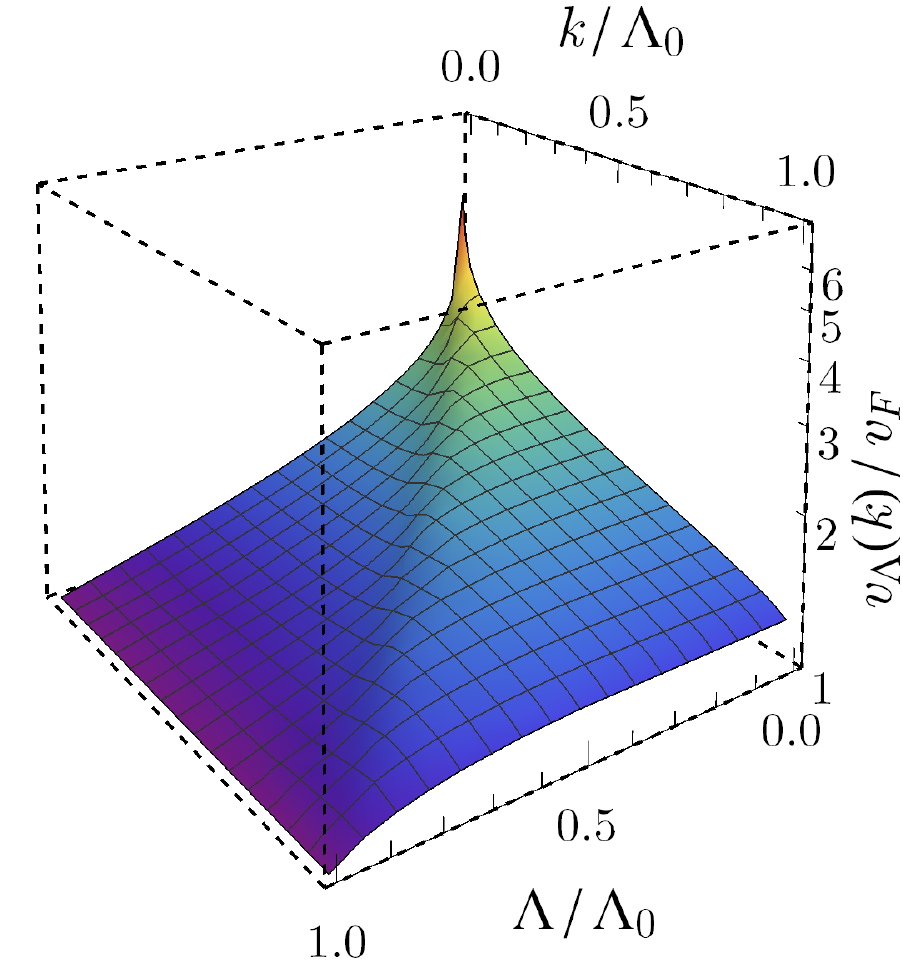}
	\caption{(Color online) Cutoff dependent Fermi velocity $v_{\Lambda}(k)$ at temperature $T/v_F \Lambda_0 = 10^{-3}$. The physical limit corresponds to $\Lambda=0$. Note that the renormalized Fermi velocity is finite at $\Lambda = k = 0$. Further observe that the figure is almost symmetric around $k = \Lambda$, suggesting that the momentum $k$ acts as an infrared cutoff for the Fermi velocity viewed as a function of $\Lambda$ in the same way as $\Lambda$ acts as a cutoff for the Fermi velocity as a function of $k$.}
	\label{fig:v-Lambda-k}
\end{figure}

\begin{figure}
	\centering
		\includegraphics[width=.48\textwidth]{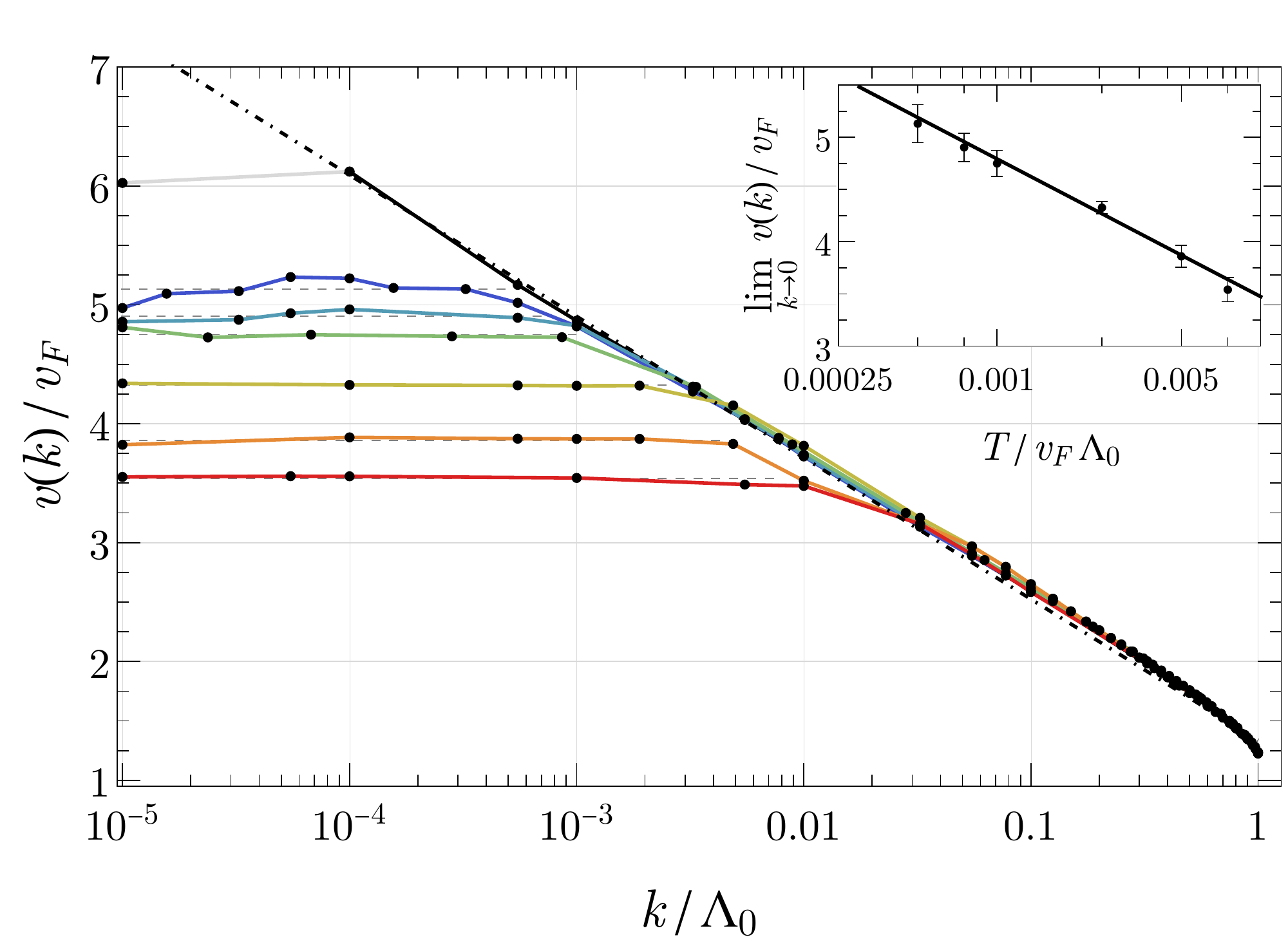}
	\caption{(Color online) Fermi velocity versus momentum $k$, for temperatures $T/v_F \Lambda_0 = 0$; $5.0 \times 10^{-4}$; $7.5 \times 10^{-4}$; $1.0 \times 10^{-3}$; $2.5 \times 10^{-3}$; $5.0 \times 10^{-3}$; and $7.5 \times 10^{-3}$ (top to bottom data sets). The inset shows the logarithmic temperature dependence of the Fermi velocity at $k = 0$. The single data point at $v(10^{-5})/v_F = 6$ shows a non-physical deviation from the logarithmic divergence at zero temperature, indicating that our numerical algorithm breaks down there. This behaviour could be expected, since the grids have only been built down to $k/\Lambda_0 = 10^{-5}$. At finite temperatures similar convergence problems occur upon approaching the lower grid cutoff $T/v_F \Lambda \approx 10^{-5}$.}
	\label{fig:FermiVelocityPhysicalLimit}
\end{figure}

\begin{figure}
	\centering
		\includegraphics[width=0.45\textwidth]{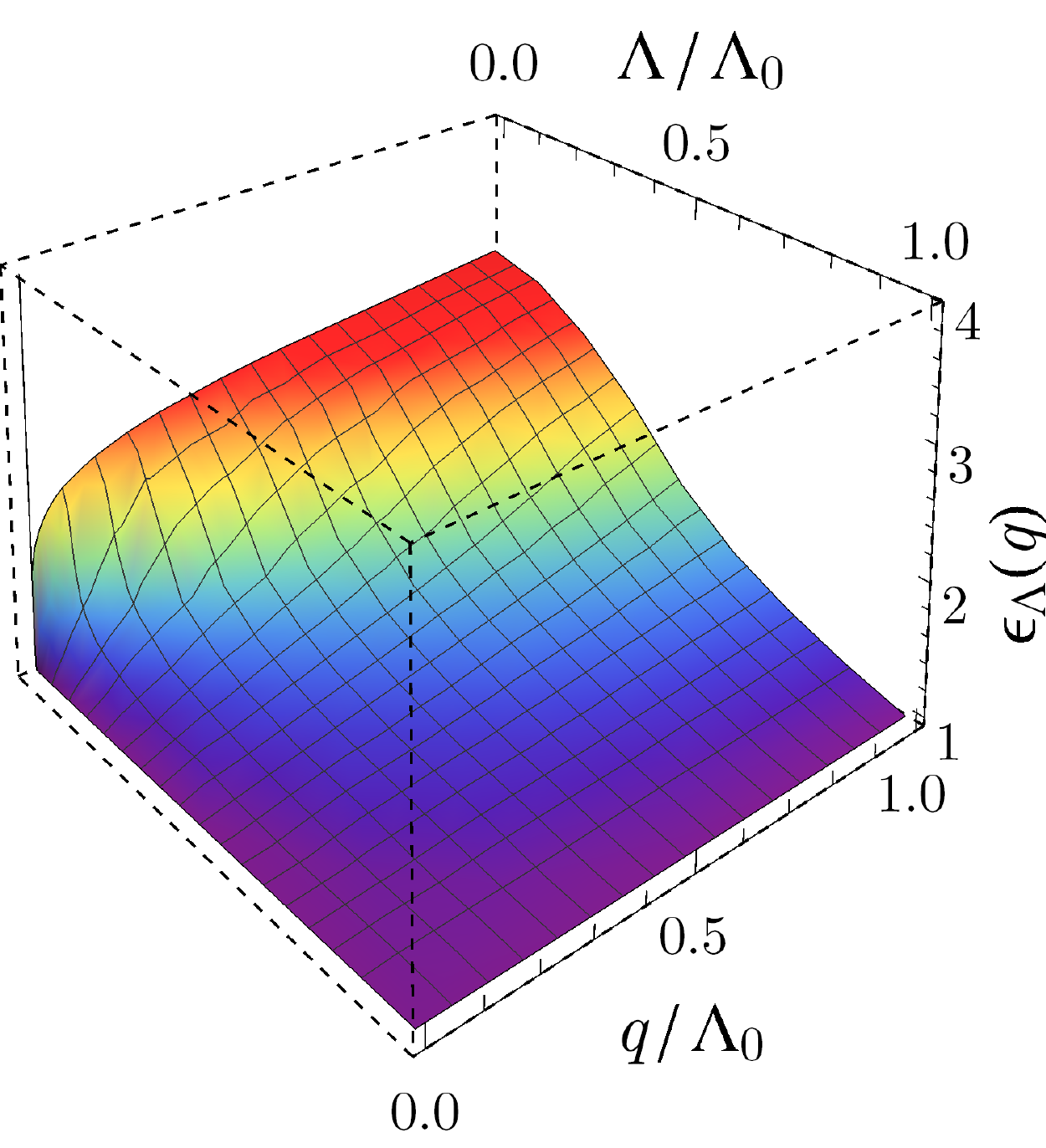}
	\caption{(Color online) Cutoff dependent dielectric function $\epsilon_{\Lambda}(q)$ at temperature $T/v_{\rm F} \Lambda_0 = 10^{-3}$. Note the sharp feature at $\Lambda = 0$ for momenta $q \lesssim T/v_{\rm F}$.}
	\label{fig:Eps-Lambda-k}
\end{figure}

\begin{figure}
	\centering
		\includegraphics[width=.48\textwidth]{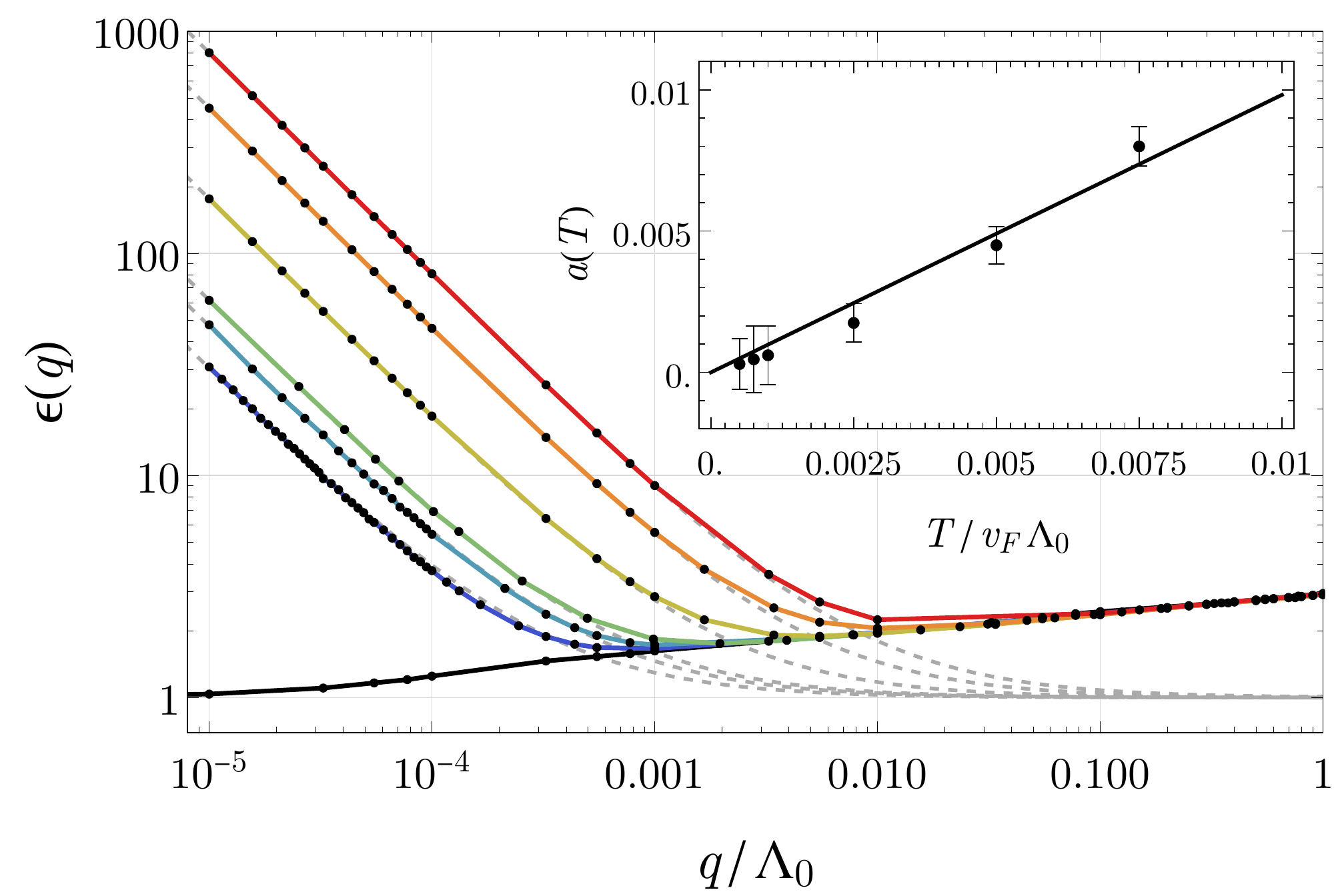}
	\caption{(Color online) Dielectric function as a function of momentum $q$ for temperatures  $T/v_{\rm F} \Lambda_0 = 0$, $5.0 \times 10^{-4}$, $7.5 \times 10^{-4}$, $1.0 \times 10^{-3}$, $2.5 \times 10^{-3}$, $5.0 \times 10^{-3}$, and $7.5 \times 10^{-3}$ (bottom to top data sets). For momenta $q/\Lambda_0$ below the reduced temperature $\tilde T = T/v_{\rm F} \Lambda_0$ our data are consistent with the $1/q$ dependence predicted by perturbation theory. The temperature dependence of the prefactor could be fitted by $\epsilon(q) = 1 + a(\tilde T) \Lambda_0/q$, indicated by dashed lines, where $a(\tilde{T})$ is a linear function as shown in the inset.}
	\label{fig:DielectricFunctionPhysicalLimit}
\end{figure}

The results of the numerical integration for the Fermi velocity $v_{\Lambda}(k)$ in its full parameter space is shown exemplarily for the reduced temperature $T/v_F \Lambda_0 = 10^{-3}$ in the figure~\ref{fig:v-Lambda-k}, whereas figure~\ref{fig:FermiVelocityPhysicalLimit} summarizes our result in the physical limit $\Lambda = 0$ for all temperatures we considered. The corresponding results for the dielectric function $\epsilon_{\Lambda}(q)$ are shown in the figures~\ref{fig:Eps-Lambda-k} and~\ref{fig:DielectricFunctionPhysicalLimit}, respectively.
%


At zero temperature the Fermi velocity shows the well-known logarithmic renormalization, which has been reported previously by many authors within one-loop perturbation theory.\cite{Vozmediano2011, CastroNetoGuineaNovoselovGeim2009, KotovEtal2012} Our numerical result could be fitted by
\begin{equation}
v(k) = A + B \, \textrm{ln} (\Lambda_0/k) \,,
\label{eq:FitVelocity}
\end{equation}
with $A = 1.34(4)$ and $B = 0.52(1)$, which coincides with the result of Bauer \textit{et al.}\cite{BauerKopietz2015} within numerical accuracy.  At nonzero temperature we find that $v$ is finite for $k \rightarrow 0$, while for large momenta the Fermi velocity merges into the logarithmic behaviour found at zero temperature. This fact can be readily explained by the presence of thermally excited charge carriers, which can screen the bare Coulomb interaction at long wavelengths. Thus, the effective Coulomb interaction becomes short ranged, cutting off the divergence at small momenta. The larger the temperature the more charge carriers are excited, leading to an enhancement in the suppression of the divergence. Indeed our numerics show that this suppression is a logarithmic function of the temperature, which could be fitted by
\begin{equation}
\lim_{k \to 0} v(k) = C + D \, \ln (v_F \Lambda_0/T) \,,
\end{equation}
with $C = 0.84(33)$ and $D = 0.57(6)$. For momenta $k \gg T/v_F$ the long-wavelength screening of the Coulomb interaction becomes irrelevant and the Fermi velocity asymptotically approaches the zero-temperature value. 

A well known issue in the comparison with experimental data is the value of the ultraviolet cutoff $\Lambda_0$. Since we already fixed the numerical value of the bare Fermi velocity by setting $\alpha = 2.2$, the cut-off $\Lambda_0$ can be used as a fit parameter. Alternatively, one could take the ultraviolet cutoff to be fixed (given by the inverse lattice spacing), and instead use $\alpha$, \textit{i.e.} $v_F$, as a fit parameter. The drawback of the latter method, however, is that the dimension of the free parameter space would be enlarged. One would have to solve the flow equations for different temperatures \textit{and} couplings $\alpha$, which would increase the numerical effort even further.

\pagebreak

The zero-temperature result for the dielectic function $\epsilon(q)$ is only very weakly momentum dependent for large momenta, while for $q \rightarrow 0$ it logarithmically approaches unity, in contrast to the momentum independence of the one-loop prediction. This behaviour is in accord with the result of Bauer \textit{et al.},\cite{BauerKopietz2015} although we observe a systematic deviation to slightly larger values at momenta of order unity. This fact may be explained by differences in the numerical implementation of the flow equations. At finite temperature, however, a strong temperature dependence, proportional to $1/q$, sets in for momenta $q \lesssim T/v_F$. The emergence of the power law divergence for small momenta can be easily understood from perturbation theory, already at the one-loop level.\cite{HwangDasSarma2007, WunschGuinea2006, SchuettGornyiMirlin2011} In the regime $q \ll T/v_F$ the static polarization function becomes momentum independent, scaling linearly with temperature, which results in the one-loop dielectric function
\begin{equation}
\epsilon_{\textrm{1-loop}}(q) = 1 + a(T) \frac{\Lambda_0}{q} \,, \quad v_F q \ll T \,,
\end{equation}
with
\begin{equation}
a(T) = 8 \, \textrm{ln} 2 \, \alpha \, \frac{T}{v_F \Lambda_0} \,.
\end{equation}
The divergence at zero momentum is a consequence of the presence of thermally excited charge carriers, screening the bare Coulomb interaction. Our numerical calculations qualitatively confirm this one-loop picture as they reproduce the $1/q$ dependence as well as the linear temperature dependence of the prefactor $a(T)$. On the quantitative level, however, we find a considerable deviation in the numerical value of the proportionality constant: The numerics could be fitted by $a(T) = 0.98(5) T/v_F \Lambda_0$, which is about one order of magnitude lower than the one-loop prediction. This discrepancy can be understood by considering the fact that a one-loop calculation employs only noninteracting propagators, while the fRG result is obtained by a selfconsistent calculation, using fully interacting propagators, such that a strong renormalization of the former result is to be expected. 

In the high-temperature limit we expect a strong screening of the Coulomb interaction, implying the absence of velocity renormalization, due to its logarithmic suppression with increasing temperatures, and hence the emergence of a free field fix point. However, such an asymptotically free fix point has little practical relevance, since in that regime the electron-phonon interaction should be taken into account in a realistic model, which would drive the system into a crumpled phase~\cite{MarianiOppen2008} and eventually lead to an instability of the underlying honeycomb lattice. In other words, graphene would have melted long before the free field fix point would have been reached.

\section{Conclusions}
\label{sec:Conclusions}
In this article we formulated a nonperturbative nonequilibrium theory for Dirac electrons interacting via the Coulomb interaction, which is based on the Keldysh functional renormalization group. Our theory should be a good description of the low-energy properties of graphene.

The essential parts of the theoretical description are the exact Dyson equations for the real-time Fermi-Bose theory, from which the quantum kinetic equations follow, as well as an exact flow equation for the effective action. The functional flow equation has been transformed into a hierarchy of ordinary coupled integro-differential equations, describing the flow of the one-particle irreducible vertex functions, by means of a vertex expansion. This hierarchy has to be solved approximately using a self-consistent truncation scheme. As a test of our formalism, we reproduced the results for the Fermi velocity renormalization and the dielectric function at zero temperature that were previously obtained by Bauer \textit{et al.}\cite{BauerKopietz2015} using the imaginary-time Matsubara formalism, and we extended these results to finite temperature.

The research provided in this article can be extended into several different ways. For equilibrium problems one may take into account dynamical effects, yielding the dynamical polarization function and quasiparticle wavefunction renormalization. This extension would go hand in hand with a nonperturbative study of collective plasmon modes. A purely bosonic cutoff combined with exact Schwinger-Dyson equations, as recently proposed by Sharma and Kopietz in Ref.~\onlinecite{SharmaKopietz2016}, would be highly advantageous for such an undertaking.

Another interesting extension is to investigate modifications of the isotropic Dirac spectrum, such as trigonal warping,\cite{CastroNetoGuineaNovoselovGeim2009} or anisotropies in strained graphene.\cite{CortijoGuineaVozmediano2015} Both phenomena would require the modification of the noninteracting Hamiltonian $H_{\textrm{f}}$, see Eq.~\eqref{eq:FreeHamiltonian}, but the general structure of the calculation is not modified. Furthermore, it would be interesting to study the fate of gaps, or masses, in the spectrum under the renormalization group flow. A particularly exciting scenario is the possibility of a \textit{spontaneous} mass generation,\cite{KotovEtal2012} for which one starts from an infinitesimal mass term at the initial scale $\Lambda = \Lambda_0$, which may be elevated to a finite value at the end of the flow. This extension, too, requires no modifications of the general formalism, as the vertex expansion in Sec.~\ref{sec:VertexExpansion} is sufficiently general enough to cope with such situations.

The application of our formalism to extrinsic graphene requires a different cutoff scheme than the one we used here, since the presence of a finite Fermi surface is incompatible with the use of a simple ``static'' momentum cutoff in the fermionic sector. One possibility would be to modify the momentum cutoff to ``dynamically'' adapt to a continuously changing Fermi surface at each scale $\Lambda$. However, this modification would complicate the flow equations considerably and is therefore not convenient.\cite{MetznerHonercampEtal2012} An alternative cutoff scheme, circumventing this difficulty, is the causality preserving frequency cutoff of Jakobs \textit{et~al.}~\cite{JakobsPletyukovSchoeller2010, JakobsMedenSchoeller2007}, which may be used either within the simple rotation invariant conical Dirac spectrum considered in this work or within one of the modifications of the bare spectrum mentioned above. Moreover, as explained at the end of Sec.~\ref{sec:FluctuationDissipationTheoremAndCutoffSchemes}, frequency cutoffs are advantageous for the study of external magnetic fields in the (integer) quantum Hall regime, since then momentum is not a well-defined quantum number, and hence cannot be employed as a flow parameter.

Whereas the use of the Keldysh formulation is technically convenient (but not essential) for equilibrium problems, because it avoids the necessity of an analytical continuation, for nonequilibrium problems the Keldysh formalism is essential. Possible applications of the formalism developed here are nonthermal fixed points, thermalization, and quantum transport in linear or even beyond linear response. Another issue of interest is the topic of nongaussian initial correlations, for which we outlined their implementation within our theoretical framework, although an actual application is beyond the scope of the present article.

For applications to realistic graphene samples, not only interactions, but also disorder has to be taken into account. This applies to quantum transport problems in particular, see Refs.~\onlinecite{SchwieteFinkelstein2014-1, SchwieteFinkelstein2014-2, KamenevBook}. The Keldysh formulation we presented here is perfectly suited for such a research programme. As is well-known the normalization of the partition function can be exploited to perform the impurity average directly on the level of the partition function. There is no need for supersymmetry or the replica trick in the Matsubara formalism. For gaussian correlated disorder the averaging procedure leads to a quartic fermionic pseudo-interaction term. Especially at the charge neutrality point the deviations from the usual Fermi-liquid behaviour should be strongly pronounced. Similarly to the Coulomb interaction treated here, the theory at this point lacks a smallness parameter and conventional approximation strategies, such as the self-consistent Born approximation, break down. Since the common truncation strategies of the infinite hierarchy of flow equations do not rely on the existence of a smallness parameter, the Keldysh fRG offers the necessary tools to go beyond these approximations in consistent manner. 

As a closing remark we want to point out that the good agreement between the functional forms of the momentum and temperature dependences of the one-loop perturbation theory and the functional renormalization group results presented here may come as a surprise, since there is no small parameter justifying a perturbative approach. Indeed, a two-loop calculation for the Fermi velocity already shows the lack of convergence of the perturbative approach, as it predicts a logarithmic decrease for small momenta.\cite{VafekCase2008} Nevertheless, the exact flow equation~\eqref{eq:ExactFlowEquationForGammaFinal} has a one-loop structure, so it becomes clear that some features of one-loop perturbation theory are qualitatively reproduced. For the future applications discussed above it is, therefore, reasonable to expect that the results derived from a perturbative one-loop calculation at least hint into the right direction, although not all features of the exact theory are reproduced quantitatively correctly. In the end, quantitatively accurate results can be expected only by more sophisticated nonperturbative approaches, such as the functional renormalization group developed here. 

\acknowledgements
We want to thank Piet Brouwer for support in the preparation of the manuscript and for discussions, as well as Severin Jakobs, Peter Kopietz, Johannes Reuter and Georg Schwiete for helpful discussions. This work is supported by the German Research Foundation (DFG) in the framework of the Priority Program 1459 ``Graphene''.

\appendix

\section{Analytical form of the vertex flow equations}
\label{sec:AnalyticalFormOfTheVertexFlowEquations}
In this appendix we give the explicit analytical form of the flow equations for the fermionic self-energy, bosonic polarization and the three-legged Fermi-Bose vertex. For completeness sake we also state the flow equation for the bosonic expectation value here again. We emphasize once more that the purely bosonic three-vertex, as well as all higher order vertices have already been neglected.

We employ here a condensed notation, where the numerical indices such as $1$ and $1'$ represent space-time coordinates $x_1 = (\vec{r}_1, t_1)$ and $x_1' = (\vec{r}_1', t_1')$, respectively, and the integration sign with a prime denotes integration over all primed space-time coordinates. Besides, the three-vertices are written in the compact form $\Gamma_{\Lambda}^{(2,1)}(1 i \alpha, 2 j \beta; 3 \gamma) = \Gamma_{\Lambda, ij}^{\alpha \beta \gamma}(1, 2; 3)$. As explained in section \ref{sec:VertexExpansion}, latin indices denote the internal degrees of freedom of the fermions (sublattice, valley, spin), and greek letters denote the Keldysh degrees of freedom (classical and quantum). In the following we also omit the $\Lambda$-indices for brevity, since all quantities appearing here are scale dependent (except for the bare Coulomb interaction~$V$), and thus there can be no confusion.

\begin{widetext}
\paragraph{Field expectation value}
\begin{align}
\int\nolimits'\ \Big( \big( V^{-1} + \Pi^R + R_{\textrm{b}}^R \big)&(1, 1') \partial_{\Lambda} \bar{\phi}_c(1') + \partial_{\Lambda} R_{\textrm{b}}^R(1, 1') \bar{\phi}_c(1') \Big) \nonumber \\
&= - \frac{i}{2} \partial \!\!\!/_{\Lambda} \sum_{\alpha, \beta} \sum_{k, l} \int\nolimits'\ G_{kl}^{\alpha \beta}(1', 2') \Gamma_{lk}^{\beta \alpha q}(2', 1'; 1) - \partial_{\Lambda} \tilde{n}(1)
\label{eq:FlowEqFieldExpectationValueAppendix}
\end{align}

\paragraph{Self-energy}
\begin{align}
\partial_{\Lambda} \Sigma_{ij}^{\alpha \beta}(1, 2) = &\int\nolimits'\ \Gamma_{ij}^{\alpha \beta c}(1, 2; 1') \partial_{\Lambda} \bar{\phi}_c(1') \nonumber \\
&+ i \partial \!\!\!/_{\Lambda} \sum_{\gamma_1, \gamma_2, \gamma_3, \gamma_4} \sum_{k, l} \int\nolimits'\ \Gamma_{ik}^{\alpha \gamma_1 \gamma_4}(1, 1'; 4') G_{kl}^{\gamma_1 \gamma_2}(1', 2') \Gamma_{lj}^{\gamma_2 \beta \gamma_3}(2', 2; 3') D^{\gamma_3 \gamma_4}(3', 4')
\label{eq:FlowEqSelfenergy}
\end{align}

\paragraph{Polarization}
\begin{align}
\partial_{\Lambda} \Pi^{\alpha \beta}(1, 2) = \frac{i}{2} \partial \!\!\!/_{\Lambda} \sum_{\gamma_1, \gamma_2, \gamma_3, \gamma_4} \sum_{k, l, m, n} \int\nolimits'\ G_{kl}^{\gamma_1, \gamma_2}(1', 2') \Gamma_{lm}^{\gamma_2 \gamma_3 \alpha}(2', 3'; 1) G_{mn}^{\gamma_3, \gamma_4}(3', 4') \Gamma_{nk}^{\gamma_4, \gamma_2 \beta}(4', 1'; 2)
\label{eq:FlowEqPolarization}
\end{align}

\paragraph{3-vertex}
\begin{align}
\partial_{\Lambda} \Gamma_{ij}^{\alpha \beta \gamma}(1, 2; 3) = i \partial \!\!\!/_{\Lambda} \sum_{\substack{\gamma_i \\ i = 1, \ldots, 6}} \sum_{k, l, m, n} \int\nolimits'\ &\Gamma_{ik}^{\alpha \gamma_1 \gamma_6}(1, 1'; 6') \Gamma_{nj}^{\gamma_4 \gamma_2 \gamma_5}(4', 2; 5') \Gamma_{lm}^{\gamma_2 \gamma_3 \gamma_3}(2', 3'; 3) \nonumber \\
&\times G_{kl}^{\gamma_1 \gamma_2}(1', 2') G_{mn}^{\gamma_3 \gamma_4}(3', 4') D^{\gamma_5 \gamma_6}(5', 6')
\label{eq:FlowEqVertex}
\end{align}

Recall that in the bare action only four of the above 3-vertices are present, namely the ones with the Keldysh indices $cqc, qcc, ccq, qqq$ being all equal to unity. The remaining three ones, with the Keldysh indices $qqc, cqq, qcq$, are generated during the flow, while the $ccc$-vertex is constrained to vanish at all scales. Therefore, we state in the following a further truncated set of the above equations, where only the four 3-vertices present in the bare action have been kept. These equations were the starting point for our analysis of thermal equilibrium in section~\ref{sec:ThermalEquilibrium}.
\\
\paragraph{Field expectation value}
\begin{align}
\int\nolimits'\ \Big( \big( V^{-1} + \Pi^R + R_{\textrm{b}}^R \big)&(1, 1') \partial_{\Lambda} \bar{\phi}_c(1') + \partial_{\Lambda} R_{\textrm{b}}^R(1, 1') \bar{\phi}_c(1') \Big) \nonumber \\
&= - \frac{i}{2} \partial \!\!\!/_{\Lambda} \sum_{k, l} \int\nolimits'\ G_{kl}^K(1', 2') \Gamma_{lk}^{ccq}(2', 1'; 1) - \partial_{\Lambda} \tilde{n}(1)
\label{eq:FlowEqFieldExpectationValueTruncatedAppendix}
\end{align}

\paragraph{Self-energy}
\begin{align}
\partial_{\Lambda} \Sigma_{ij}^R(1, 2) = \int\nolimits'\ \Gamma_{ij}^{qcc}(1, 2; 1') \partial_{\Lambda} \bar{\phi}_c(1') + i \partial \!\!\!/_{\Lambda} \sum_{k, l} \int\nolimits'\ \Big( &\Gamma_{ik}^{qcc}(1, 1'; 4') G_{kl}^K(1', 2') \Gamma_{lj}^{ccq}(2', 2; 3') D^A(3', 4') \nonumber \\ 
+ &\Gamma_{ik}^{qcc}(1, 1'; 4') G_{kl}^R(1', 2') \Gamma_{lj}^{qcc}(2', 2; 3') D^K(3', 4') \Big)
\label{eq:FlowEqRetardedSelfenergy}
\end{align}
\begin{align}
\partial_{\Lambda} \Sigma_{ij}^A(1, 2) = \int\nolimits'\ \Gamma_{ij}^{cqc}(1, 2; 1') \partial_{\Lambda} \bar{\phi}_c(1') + i \partial \!\!\!/_{\Lambda} \sum_{k, l} \int\nolimits'\ \Big( &\Gamma_{ik}^{ccq}(1, 1'; 4') G_{kl}^K(1', 2') \Gamma_{lj}^{cqc}(2', 2; 3') D^R(3', 4') \nonumber \\
+ &\Gamma_{ik}^{cqc}(1, 1'; 4') G_{kl}^A(1', 2') \Gamma_{lj}^{cqc}(2', 2; 3') D^K(3', 4') \Big)
\label{eq:FlowEqAdvancedSelfenergy}
\end{align}
\begin{align}
\partial_{\Lambda} \Sigma_{ij}^K(1, 2) = i \partial \!\!\!/_{\Lambda} \sum_{k, l} \int\nolimits'\ \Big( &\Gamma_{ik}^{qcc}(1, 1'; 4') G_{kl}^K(1', 2') \Gamma_{lj}^{cqc}(2', 2; 3') D^K(3', 4') \nonumber \\
+ &\Gamma_{ik}^{qcc}(1, 1'; 4') G_{kl}^R(1', 2') \Gamma_{lj}^{qqq}(2', 2; 3') D^A(3', 4') \nonumber \\
+ &\Gamma_{ik}^{qqq}(1, 1'; 4') G_{kl}^A(1', 2') \Gamma_{lj}^{cqc}(2', 2; 3') D^R(3', 4') \Big)
\label{eq:FlowEqKeldyshSelfenergy}
\end{align}
\begin{align}
0 \stackrel{!}{=} \partial_{\Lambda} \Sigma_{ij}^Z(1, 2) = i \partial \!\!\!/_{\Lambda} \sum_{k, l} \int\nolimits'\ \Big( &\Gamma_{ik}^{ccq}(1, 1'; 4') G_{kl}^R(1', 2') \Gamma_{lj}^{qcc}(2', 2; 3') D^R(3', 4') \nonumber \\
+ &\Gamma_{ik}^{cqc}(1, 1'; 4') G_{kl}^A(1', 2') \Gamma_{lj}^{ccq}(2', 2; 3') D^A(3', 4') \Big)
\label{eq:FlowEqZComponentSelfenergy}
\end{align}

\paragraph{Polarization}
\begin{align}
\partial_{\Lambda} \Pi^R(1, 2) = \frac{i}{2} \partial \!\!\!/_{\Lambda} \sum_{k, l, m, n} \int\nolimits'\ \Big( &G_{kl}^K(1', 2') \Gamma_{lm}^{ccq}(2', 3'; 1) G_{mn}^R(3', 4') \Gamma_{nk}^{qcc}(4', 1'; 2) \nonumber \\
+ &G_{kl}^A(1', 2') \Gamma_{lm}^{ccq}(2', 3'; 1) G_{mn}^K(3', 4') \Gamma_{nk}^{cqc}(4', 1'; 2) \Big)
\label{eq:FlowEqRetardedPolarization}
\end{align}
\begin{align}
\partial_{\Lambda} \Pi^A(1, 2) = \frac{i}{2} \partial \!\!\!/_{\Lambda} \sum_{k, l, m, n} \int\nolimits'\ \Big( &G_{kl}^K(1', 2') \Gamma_{lm}^{cqc}(2', 3'; 1) G_{mn}^A(3', 4') \Gamma_{nk}^{ccq}(4', 1'; 2) \nonumber \\
+ &G_{kl}^R(1', 2') \Gamma_{lm}^{qcc}(2', 3'; 1) G_{mn}^K(3', 4') \Gamma_{nk}^{ccq}(4', 1'; 2) \Big)
\label{eq:FlowEqAdvancedPolarization}
\end{align}
\begin{align}
\partial_{\Lambda} \Pi^K(1, 2) = \frac{i}{2} \partial \!\!\!/_{\Lambda} \sum_{k, l, m, n} \int\nolimits'\ \Big( &G_{kl}^K(1', 2') \Gamma_{lm}^{ccq}(2', 3'; 1) G_{mn}^K(3', 4') \Gamma_{nk}^{ccq}(4', 1'; 2) \nonumber \\
+ &G_{kl}^R(1', 2') \Gamma_{lm}^{qqq}(2', 3'; 1) G_{mn}^A(3', 4') \Gamma_{nk}^{ccq}(4', 1'; 2) \nonumber \\
+ &G_{kl}^A(1', 2') \Gamma_{lm}^{ccq}(2', 3'; 1) G_{mn}^R(3', 4') \Gamma_{nk}^{qqq}(4', 1'; 2) \Big)
\label{eq:FlowEqKeldyshPolarization}
\end{align}
\begin{align}
0 \stackrel{!}{=} \partial_{\Lambda} \Pi^Z(1, 2) = \frac{i}{2} \partial \!\!\!/_{\Lambda} \sum_{k, l, m, n} \int\nolimits'\ \Big( &G_{kl}^R(1', 2') \Gamma_{lm}^{qcc}(2', 3'; 1) G_{mn}^R(3', 4') \Gamma_{nk}^{qcc}(4', 1'; 2) \nonumber \\
+ &G_{kl}^A(1', 2') \Gamma_{lm}^{cqc}(2', 3'; 1) G_{mn}^A(3', 4') \Gamma_{nk}^{cqc}(4', 1'; 2) \Big)
\label{eq:FlowEqZComponentPolarization}
\end{align}

\paragraph{3-vertex}
\begin{align}
\partial_{\Lambda} \Gamma_{ij}^{ccq}(1, 2; 3) = i \partial \!\!\!/_{\Lambda} \sum_{k, l, m, n} \int\nolimits'\ \Big( &D^K(5', 6') \Gamma_{ik}^{cqc}(1, 1'; 6') G_{kl}^A(1', 2') \Gamma_{lm}^{ccq}(2', 3'; 3) G_{mn}^R(3', 4') \Gamma_{nj}^{qcc}(4', 2; 5') \nonumber \\
+ &D^R(5', 6') \Gamma_{ik}^{ccq}(1, 1'; 6') G_{kl}^K(1', 2') \Gamma_{lm}^{ccq}(2', 3'; 3) G_{mn}^R(3', 4') \Gamma_{nj}^{qcc}(4', 2; 5') \nonumber \\
+ &D^A(5', 6') \Gamma_{ik}^{cqc}(1, 1'; 6') G_{kl}^A(1', 2') \Gamma_{lm}^{ccq}(2', 3'; 3) G_{mn}^K(3', 4') \Gamma_{nj}^{ccq}(4', 2; 5') \Big)
\label{eq:FlowEqVertexCCQ}
\end{align}
\begin{align}
\partial_{\Lambda} \Gamma_{ij}^{cqc}(1, 2; 3) = i \partial \!\!\!/_{\Lambda} \sum_{k, l, m, n} \int\nolimits'\ \Big( &D^K(5', 6') \Gamma_{ik}^{cqc}(1, 1'; 6') G_{kl}^A(1', 2') \Gamma_{lm}^{cqc}(2', 3'; 3) G_{mn}^A(3', 4') \Gamma_{nj}^{cqc}(4', 2; 5') \nonumber \\
+ &D^R(5', 6') \Gamma_{ik}^{ccq}(1, 1'; 6') G_{kl}^K(1', 2') \Gamma_{lm}^{cqc}(2', 3'; 3) G_{mn}^A(3', 4') \Gamma_{nj}^{cqc}(4', 2; 5') \nonumber \\
+ &D^R(5', 6') \Gamma_{ik}^{ccq}(1, 1'; 6') G_{kl}^R(1', 2') \Gamma_{lm}^{qcc}(2', 3'; 3) G_{mn}^K(3', 4') \Gamma_{nj}^{cqc}(4', 2; 5') \Big)
\label{eq:FlowEqVertexCQC}
\end{align}
\begin{align}
\partial_{\Lambda} \Gamma_{ij}^{qcc}(1, 2; 3) = i \partial \!\!\!/_{\Lambda} \sum_{k, l, m, n} \int\nolimits'\ \Big( &D^K(5', 6') \Gamma_{ik}^{qcc}(1, 1'; 6') G_{kl}^R(1', 2') \Gamma_{lm}^{qcc}(2', 3'; 3) G_{mn}^R(3', 4') \Gamma_{nj}^{qcc}(4', 2; 5') \nonumber \\
+ &D^A(5', 6') \Gamma_{ik}^{qcc}(1, 1'; 6') G_{kl}^R(1', 2') \Gamma_{lm}^{qcc}(2', 3'; 3) G_{mn}^K(3', 4') \Gamma_{nj}^{ccq}(4', 2; 5') \nonumber \\
+ &D^A(5', 6') \Gamma_{ik}^{qcc}(1, 1'; 6') G_{kl}^K(1', 2') \Gamma_{lm}^{cqc}(2', 3'; 3) G_{mn}^A(3', 4') \Gamma_{nj}^{ccq}(4', 2; 5') \Big)
\label{eq:FlowEqVertexQCC}
\end{align}
\begin{align}
\partial_{\Lambda} \Gamma_{ij}^{qqq}(1, 2; 3) = i \partial \!\!\!/_{\Lambda} \sum_{k, l, m, n} \int\nolimits'\ \Big( &D^K(5', 6') \Gamma_{ik}^{qcc}(1, 1'; 6') G_{kl}^K(1', 2') \Gamma_{lm}^{ccq}(2', 3'; 3) G_{mn}^K(3', 4') \Gamma_{nj}^{cqc}(4', 2; 5') \nonumber \\
+ &D^K(5', 6') \Gamma_{ik}^{qcc}(1, 1'; 6') G_{kl}^R(1', 2') \Gamma_{lm}^{qqq}(2', 3'; 3) G_{mn}^A(3', 4') \Gamma_{nj}^{cqc}(4', 2; 5') \nonumber \\
+ &D^R(5', 6') \Gamma_{ik}^{qqq}(1, 1'; 6') G_{kl}^A(1', 2') \Gamma_{lm}^{ccq}(2', 3'; 3) G_{mn}^K(3', 4') \Gamma_{nj}^{cqc}(4', 2; 5') \nonumber \\
+ &D^A(5', 6') \Gamma_{ik}^{qcc}(1, 1'; 6') G_{kl}^K(1', 2') \Gamma_{lm}^{ccq}(2', 3'; 3) G_{mn}^R(3', 4') \Gamma_{nj}^{qqq}(4', 2; 5') \Big)
\label{eq:FlowEqVertexQQQ}
\end{align}
\end{widetext}

\end{fmffile}
\end{document}